\documentclass[11pt]{article}
\pdfoutput=1
\usepackage{aas_macros,amsmath,amssymb,comment,cite,esint,graphicx,mathtools}
\usepackage[margin=.8in,letterpaper]{geometry}
\usepackage[colorlinks=true]{hyperref}
\usepackage[affil-it]{authblk}
\usepackage{url}
\usepackage{ulem}
\usepackage{overpic}

\numberwithin{equation}{section}
\let\originalleft\left
\let\originalright\right
\renewcommand{\left}{\mathopen{}\mathclose\bgroup\originalleft}
\renewcommand{\right}{\aftergroup\egroup\originalright}
\mathcode`\*="8000
{\catcode`\*=\active\gdef*{\mathclose{}\,\mathopen{}}}

\newcommand{\br}[1]{\left[#1\right]}

\newcommand{\pa}[1]{\left(#1\right)}
\newcommand{\td}[1]{\tilde{#1}}
\newcommand{\M}[1]{\mathcal{#1}}
\newcommand{\isco}{\text{ISCO}}

\newcommand{\mob}{\text{mob}}

\newcommand{\be}{\begin{equation}}
\newcommand{\ee}{\end{equation}}
\newcommand{\bea}{\setlength\arraycolsep{2pt} \begin{eqnarray}}
\newcommand{\eea}{\end{eqnarray}}

\def \a {\alpha}
\def \b {\beta}
\def \g {\gamma}

\def \d {\delta}
\def \D {\Delta}
\def \e {\epsilon}

\def \m {\mu}

\def \k {\kappa}
\def \l {\lambda}

\def \s {\sigma}
\def \S {\Sigma}

\def \th {\theta}
\def \Th {\Theta}
\def \t {\tau}

\def \p {\partial}
\def \f {\frac}

\begin{document}
\title{Photon emissions from Kerr equatorial geodesic orbits}

\author{
Yanming Su$^{1}$, Minyong Guo$^{2}$, Haopeng Yan$^{3\ast}$,
Bin Chen$^{1,4,5}$}
\date{}

\maketitle

\vspace{-10mm}

\begin{center}
{\it
$^1$Department of Physics, Peking University, No.5 Yiheyuan Rd, Beijing
100871, People's Republic of China\\\vspace{4mm}

$^2$ Department of Physics, Beijing Normal University,
Beijing 100875, People's Republic of China\\\vspace{4mm}

$^3$ College of Physics, Taiyuan University of Technology, Taiyuan, 030024, People's Republic of China\\\vspace{4mm}

$^4$ Center for High Energy Physics, Peking University,
No.5 Yiheyuan Rd, Beijing 100871, People's Republic of China\\\vspace{4mm}

$^5$ Collaborative Innovation Center of Quantum Matter,
No.5 Yiheyuan Rd, Beijing 100871, People's Republic of China\\\vspace{2mm}
}
\end{center}

\vspace{8mm}

\begin{abstract}
We consider the light emitters moving freely along the geodesics on the equatorial plane near a Kerr black hole and study the observability of these emitters. To do so, we assume these emitters emit photons isotropically and monochromatically, and we compute the photon escaping probability (PEP) and the maximum observable blueshift (MOB) of the photons that reach infinity. We obtain numerical results of PEP and MOB for the emitters along various geodesic orbits, which exhibit distinct features for the trajectories of different classes. In particular, we find that the plunging emitters could have considerable observability even in the near-horizon region. This interesting observational feature becomes more significant for the high-energy emitters near a high-spin black hole. As the radiatively-inefficient accretion flow may consist of plunging emitters, the present work could be of great relevance to the astrophysical observations.
\end{abstract}

\vfill{\footnotesize $^\ast$ Corresponding author: yanhaopeng@tyut.edu.cn}

\maketitle

\newpage

\section{Introduction}\label{sec:Introduction}
In recent years, the Event Horizon Telescope (EHT) Collaboration has released the images of the supermassive black holes M87* and Sgr A* in succession \cite{EventHorizonTelescope:2019dse, EventHorizonTelescope:2022xnr}. A bright ring and a central dark shadow have been observed in each EHT image. The colorful rings are asymmetric in brightness and are composed of the photons emitted from the equatorial accretion disks, and the dark shadows correspond to the regions where the black holes capture the photons. It is a remarkable progress in observing a black hole at the event horizon scale, which attracts broad research interests on many observational aspects of black holes, including, for example, the shadows \cite{Cunha:2018acu, Perlick:2021aok, Wei:2019pjf,Banerjee:2022jog,Banerjee:2019nnj,Mishra:2019trb,Li:2021btf,
Wang:2022kvg,Li:2020drn,Zhong:2021mty}, the photon rings \cite{EventHorizonTelescope:2021bee,Gralla:2019xty,Himwich:2020msm,Johnson:2019ljv,Gralla:2020srx,
Peng:2020wun,Peng:2021osd,Chen:2022nbb},  the signatures of surrounding hot spots \cite{Gralla:2017ufe,Guo:2018kis,Yan:2019etp,Guo:2019lur} and the magnetospheres\cite{EventHorizonTelescope:2021srq,EventHorizonTelescope:2021btj,
Junior:2021dyw,Hou:2022eev,Hu:2022sej,Lee:2022rtg}. The photon ring's brightness and size also encode the information of the accretion disk. Therefore, it is essential to study the photons emissions from the particles in the accretion disk, and answer the following question: how many photons can escape to infinity, and how are the photon frequencies shifted at infinity?

The escaping of the photons in the Schwarzschild spacetime was first studied by Synge in \cite{Synge:1966okc}, where it was found that the escape cone shrinks as the emitting point shifts towards the horizon. Later, the ``photon escape cone'' in the Kerr spacetime was studied by Semerak in \cite{etde_271389}, where both the black hole and the naked-singularity cases were discussed.
The ``light escape cone'' in the Kerr-de Sitter spacetime was studied in \cite{Stuchlik:2018qyz}, where the sources in both radial geodesic and circular geodesic motions were considered.
Recently, the photon escaping from the light sources of different motions had been extensively studied. The photon escaping probability (PEP) of zero-angular momentum sources (ZAMS) was first studied for the Kerr-Newman spacetime in \cite{Ogasawara:2019mir}, where the extremal, sub-extremal and non-extremal cases were discussed separately. Soon after, the PEP of ZAMS for the Kerr-Sen spacetime was studied in \cite{Zhang:2020pay}. The PEP and maximum observable blueshift (MOB) of the photon emissions from the sources on circular geodesics outside the ISCO of a Kerr black hole were studied in \cite{Igata:2019hkz, Gates:2020els}. The PEP and MOB of the plunging emitters starting from the ISCO of a Kerr black hole were studied in \cite{Igata:2021njn}. It was found that the PEP of a ZAMS approaching the event horizon tends to zero and $29\%$ for a non-extremal and extremal Kerr black hole, respectively \cite{Ogasawara:2019mir}, and the PEP of circular emitters at the ISCO is larger than $55\%$ \cite{Igata:2019hkz, Gates:2020els}, while the PEP from plunging emitters at approximately halfway between the ISCO and horizon is about $50\%$ \cite{Igata:2021njn}. Moreover, it was also found that the emitter's proper motion affects the MOB of the escaped photons \cite{Igata:2019hkz, Gates:2020els, Igata:2021njn}. Very recently, the condition for the photons escaping from the off-equator sources to the infinity in the Kerr spacetime was clarified in \cite{Ogasawara:2020frt, Ogasawara:2021yfe}.

For extremal and near-extremal Kerr black holes, the escaping of the photons could be investigated more clearly in the near-horizon extremal Kerr (NHEK) and near-horizon near-extremal Kerr (near-NHEK) geometries. By using the (near-)NHEK\footnote{We use ``(near-)NHEK'' to represent both NHEK and near-NHEK.} metrics \cite{Bardeen:1999px, Bredberg:2009pv, Gralla:2015rpa}, the calculations of PEP and MOB were simplified, and some analytical results were obtained \cite{Gates:2020els, Yan:2021yuo, Yan:2021ygy}. The PEP and the blueshift distributions of the emitters moving at the ISCO (residing in the NHEK region) were reproduced analytically in \cite{Gates:2020els}. Following the method of \cite{Gates:2020els}, the photon emissions from ZAMS in both the NHEK and near-NHEK regions were analytically studied in \cite{Yan:2021yuo}. Then in \cite{Yan:2021ygy}, the photon emissions from equatorial emitters following various geodesic motions in the (near-)NHEK geometry were further studied. It was found that the PEP for ZAMS in the NHEK region and for the source at the innermost photon orbit in the near-NHEK region  is  about $29\%$ and $13\%$ respectively; the PEP  is larger than $50\%$ for the outgoing geodesic emitters that can eventually reach NHEK infinity; the PEP   is less than $55\%$ for the plunging geodesic emitters that ultimately enters the horizon, and the PEP is less than $59\%$ for the bounded geodesic emitters in the (near-)NHEK region. It was also found that all escaping photons from ZAMS are redshifted due to the strong gravity, while those escaping photons from the emitters with various motions could be blueshifted when the Doppler effect overwhelms the strong gravity effect.

So far, the escaping of the photons from generic light sources for general non-extremal Kerr black holes has not been studied.
This paper will study the photon emissions from the equatorial sources along all the possible geodesic orbits in the Kerr exterior, including generic plunging orbits, trapped orbits, bounded orbits, and deflected orbits.
This work generalizes the results in \cite{Igata:2021njn}, where only the marginal plunging orbits from the ISCO were considered. On the other hand, this work also generalizes the previous studies for the (near-)NHEK cases \cite{Yan:2021ygy} to the general-spin Kerr case.

The remaining part of this paper is organized as follows. In Sec.~\ref{sec:Geodesics}, we review the timelike and null geodesics in the Kerr spacetime and then introduce a classification of the equatorial timelike geodesics. In Sec.~\ref{sec:photonescapes}, we discuss the problem of photon emissions from an equatorial emitter and obtain the formulae of the PEP and MOB. In Sec.~\ref{sec:PEPandMOB}, we display the numerical results for the PEP and MOB by using the figures and the tables and discuss them in detail. In Sec.~\ref{sec:conclusion}, we conclude this work.

\section{Geodesics in the Kerr exterior}\label{sec:Geodesics}
The Kerr spacetime metric in the Boyer-Lindquist coordinates $x^\m=(t,r,\th,\phi)$ is given by
\be
\label{metric1}
ds^2=-\frac{\Sigma\Delta}{\Xi}dt^2+\frac{\Sigma}{\Delta}dr^2+\Sigma d\theta^2+\frac{\Xi\sin^2\theta}{\Sigma}\pa{d\phi-\frac{2Mar}{\Xi} dt}^2,
\ee
where
\be
\label{metric2}
\Delta=r^2-2Mr+a^2,\qquad
\Sigma=r^2+a^2\cos^2\theta,\qquad
\Xi=(r^2+a^2)^2-a^2\Delta\sin^2\theta.
\ee
Here, $M$ and $a$ are the mass and spin parameters of a black hole, respectively, and the spin $a$ is defined by $a=J/M$ with $J$ being the angular momentum.
The outer event horizon of the black hole is located at
\be
\label{horizon}
r_H=M+\sqrt{M^2-a^2}.
\ee
In the following, we set $M=1$ for simplicity.

There are four conserved quantities of a free particle: the mass $\mu$, the energy $E$, the axial angular momentum $l$, and the Carter constant $Q$.
One can derive the four-momentum $p^\mu$ of this particle by using the Hamilton-Jacobi method \cite{Bardeen:1973tla}, which reads
\bea
\label{kerrPr}
p^r&=&\pm_r\f{1}{\S}\sqrt{\M{R}(r)},\\
\label{kerrPth}
p^\th&=&\pm_\th\f{1}{\S}\sqrt{\Th(\th)},\\
\label{kerrPphi}
p^\phi&=&\f{1}{\S}\br{\f{a}{\D}[E(r^2+a^2)-al]+\f{l}{\sin^2\th}-aE},\\
\label{kerrPt}
p^t&=&\f{1}{\S}\br{\f{r^2+a^2}{\D}[E(r^2+a^2)-al]+a(l-aE\sin^2\th)},
\eea
where
\bea
\label{kerrRPotential}
\M{R}(r)&=&[E(r^2+a^2)-al]^2-\D[Q+(l-aE)^2+\mu^2r^2],\\
\label{kerrThpotential}
\Th(\th)&=&Q+a^2(E^2-\mu^2)\cos^2\th-l^2\cot^2\th,
\eea
are the radial and angular potentials, respectively, and $\pm_r$ and $\pm_\th$ denote the signs of the radial and polar motions, respectively.

For a massive timelike particle, we have $\mu>0$ and $p^\mu=\mu\f{\p x^\mu}{\p \t}$ with $\t$ being the proper time.
For a massless particle (photon), we have $\mu=0$ and $p^\mu=\f{\p x^\mu}{\p\tau}$ with $\t$ being an affine parameter. For photons, it is convenient to express $p^\m$ under a reparameterization by using the two rescaled quantities:
\be
\label{impactparameters}
\l=\f{l}{E},\qquad
\eta=\f{Q}{E^2}.
\ee

We now consider the equatorial timelike geodesics in the Kerr exterior. Hereafter, we set $\mu=1$ and let $s_r=\pm_r$ for timelike particles. Then $s_r=-1$ and $s_r=+1$ are for ingoing and outgoing particles, respectively.
On the equatorial plane, we have $Q=0$, a geodesic is determined by the energy $E$ and angular momentum $l$.
By studying the root structure of the radial potential $\M R(r)$, the radial motions are classified in the $(E,l)$ phase space in \cite{Compere:2021bkk}. In the following, we present a short review of the classification.

For geodesics that can reach the horizon, the energy $E$ and angular momentum $l$ is constraint by the thermodynamic bound:
\be
\label{thermbound}
l\leq l_H(E,a)=\frac{ E}{\Omega_H}=\frac{2E}{a}(1+\sqrt{1-a^2}),
\ee
in which $\Omega_H=\f{a}{2r_H}$ is the angular velocity of the event horizon. In the ergosphere, this is also the superradiation bound. When $l=l_H$, the potential $\M R(r)$ has one root at the horizon. On the contrary, geodesics with $l>l_H$ cannot reach the horizon.

Let us consider the double root structure of $\M R(r)$. A double root corresponds to a circular orbit, and we use the subscript ``$_\ast$'' to represent a double root.
For the $l=l_H$ case, we have one root at the horizon, and the double root requires $Y(r_\ast)=Y^\prime(r_\ast)=0$ with $Y(r)=\M R(r)/(r-r_H)$. The solution is given by $E=E_c$ with
\be
E_c=\f{r_H+2\sqrt{r_H-1}}{\sqrt{r_H(r_H+2+4\sqrt{r_H-1})}}.
\ee
Such a double root corresponds to a stable circular orbit since $Y^{\prime\prime}(r_\ast)<0$.
For $l\neq l_H$ case, we have $\M R(r_\ast)=\M R^\prime(r_\ast)=0$, then the solutions are given by \cite{Bardeen:1972fi}
\bea
\label{doublertE}
&&E_\pm(r_\ast)=\frac{r_\ast(r_\ast-2)\pm a r_\ast^{1/2}}{ \sqrt{r_\ast^3(r_\ast-3)\pm 2ar_\ast^{5/2}}},\\
\label{doublertl}
&&l_\pm(r_\ast)=\pm\frac{(r_\ast^2+a^2)r_\ast^{1/2}\mp 2ar_\ast}{\sqrt{r_\ast^3(r_\ast-3)\pm 2ar_\ast^{5/2}}}.
\eea
Hereafter the plus/minus sign ``$\pm$'' represents the prograde/retrograde orbit, respectively.
The ranges of the double roots are bounded by the innermost (prograde) and outermost (retrograde) photon orbits $r_{p\pm}$, that is $r_\ast>r_{p\pm}$, where
\be
\label{photonboundary}
r_{p\pm}=2\br{1+\cos\pa{\frac{2}{3}\arccos(\mp a)}}.
\ee
Some of these circular orbits are stable, while others are unstable.
The marginal stable circular orbits have $\M R^{\prime\prime}(r_\ast)=0$, the solutions are the triple roots 
\be
\label{tripleroot}
r_\ast=r_{\isco\pm}=3+Z_2\mp\sqrt{(3-Z_1)(3+Z_1+2Z_2)},
\ee
where
\be
Z_1=1+(1-a^2)^{\f{1}{3}}\br{(1+a)^{\f{1}{3}}+(1-a)^{\f{1}{3}}},\qquad
Z_2=\sqrt{3a^2+Z_1^2}.
\ee
By substituting the triple roots \eqref{tripleroot} into Eqs.~\eqref{doublertE} and \eqref{doublertl}, we have
\bea
&&E_{\isco \pm}=E_\pm(r_{\isco\pm}),\\
&&l_{\isco \pm}=l_\pm(r_{\isco \pm}).
\eea
Inverting Eq.~\eqref{doublertE} and substituting the results into Eq.~\eqref{doublertl}, we can formally write the angular momentum $l$ as functions of the energy $E$, that is
\bea
&&l^u(E)=l^{u}_\pm(E)=l^{u}_\pm(r_\ast^u(E)),\\
&&l^s(E)=l^{s}_\pm(E)=l^{s}_\pm(r_\ast^s(E)).
\eea
Hereafter, we use the superscripts ``$u$'' and ``$s$'' to denote the unstable and stable double roots.
Note that we have $l=l^u$ for $r_{p\pm}<r_\ast<r_{\isco\pm}$ and we have $l=l^s$ for $r_{\isco\pm}<r_\ast$.

Based on the radial root structures, the equatorial motions are classified in the $(E,l)$ phase space, as shown in Table~\ref{table:classification}. First, we introduce the four basic classes of geodesic orbits which only involve single roots: the plunging orbits $\M P$, the trapped orbits $\M T$, the deflected orbits $\M D$, and the bounded orbits $\M B$. A plunging particle plunges into the black hole from infinity. A trapped particle emerges from the white hole and falls into the black hole after bouncing back from a turning point. A deflected particle comes from infinity and bounces back from a turning point. A bounded particle oscillates between two turning points. Note that an anti-plunging (anti$-\M P$) particle travels on the same trajectory as its counter partner but has the opposite radial orientation. Next, we consider the classes of orbits that involve unstable double roots, that is, $l=l^u(E)$. We call such orbits the marginal orbits. When a particle travels across the unstable double root on a marginal orbit\footnote{It takes infinite amount of proper time for a marginal particle to approach the unstable double root, so that the ``marginal orbits'' are asymptotic orbits for the emitters having near-critical parameters, and they are not physical orbits.}, it either plunges toward the horizon or moves toward infinity. Depending on whether a particle could or could not potentially reach infinity, the relevant orbits are named the marginal plunging orbits $\M {MP}$ or marginal trapped orbits $\M {MT}$, respectively\footnote{A more accurate and detailed classification for these marginal geodesic motions can be found in \cite{Compere:2021bkk}, here we introduce this simplified version of classification for convenience.}. Besides, the orbit involves a triple root named the marginally trapped orbit from the ISCO $\M{MT}_\isco$. At a double root or a triple root, the relevant circular orbits have three classes: the stable circular orbit $\M C^s$, the unstable circular orbit $\M C^u$ and the ISCO $\M C_\isco$.
\begin{table}[h]
\centering
\caption{Classification of equatorial timelike geodesics (see Fig.~8 and Table V of \cite{Compere:2021bkk} for a more sophisticated classification). We follow the notations in \cite{Compere:2021bkk} for the root structures: the symbols $\vert$, $+$, $-$ and $\rangle$ represent the outer horizon $r_H$, a region where $\M R (r)>0$, a region where $\M R (r)<0$ and the radial infinity, respectively; the symbols $\bullet$, $\bullet\hspace{-2pt}\bullet$, $\bullet\hspace{-4pt}\bullet\hspace{-4pt}\bullet$ and $\vert\hspace{-4.5pt}\bullet$ denote a single root, a double root, a triple root and a root at the outer horizon, respectively. In addition, we also use dots $\dots$ to cover all possible root structures. We label the roots with $r_1,~r_2$ and $r_3$ ($r_1<r_2<r_3$), and we also use $r_{\text{last}}$ to denote the last root for the deflected case $\M D$.
}
\begin{tabular}{c c c c c }
\hline \hline
Name & Root structure & Energy range & Angular momentum range & Radial range \\
\hline
(anti-)$\M P$ & $ \vert+\rangle$ & $E\geq 1$ & $l^{u}_-<l<l^{u}_+$ & $r_+\leq r< \infty$\\
\hline
$\M T$ & $ \vert+\bullet -\dots $ & $ E<1$ & $l<l_H$ & $r_+\leq r\leq r_1$ \\
\hline
$\M D$ & $\dots- \bullet+\rangle$ & $E\geq1$ & $l<l^{u}_-$ or $l>l^{u}_+$, & $r_{\text{last}}\leq r<\infty$\\
\hline
$\M B$ & $\vert +\bullet - \bullet + \bullet -\rangle$ & $E_{\isco-}<E<1$
& $l^{s}_-<l<l^{u}_-<0,$ & $r_2\leq r\leq r_3$\\
& & $E_{\isco+}<E<1$ & $0<l^{u}_+<l<\text{min}(l_H,l^{s}_+)$ & \\
& $\vert\hspace{-6.6pt}\bullet- \bullet+\bullet-\rangle$ & $E_c<E<1$
& $l=l_H$ &$r_2\leq r\leq r_3$\\
\hline
(anti-)$\M{MP}$ & $ \vert+\bullet\hspace{-4pt}\bullet\hspace{2pt}+ \rangle$ & $E\geq1$ & $l=l^u$ & $r_+\leq r< \infty$\\
\hline
$\M {MT}$ & $ \vert+\bullet\hspace{-4pt}\bullet\hspace{2pt}+\bullet -\rangle$ &
$E_{\isco\pm}<E<1$ & $l=l^{u}$ & $r_+\leq r\leq r_2$ \\
\hline
$\M{MT}_\isco$ & $ \vert+ \bullet\hspace{-4pt}\bullet\hspace{-4pt}\bullet\hspace{2pt}-\rangle$
& $ E= E_{\isco \pm}$ & $ l= l_{\isco \pm}$ & $r_+\leq r<r_\isco$\\
\hline
$\M C^s$ & $\dots- \bullet\hspace{-4pt} \bullet\hspace{2pt}-\rangle$ & $E_{\isco\pm}<E<1$ & $l=l^s$ & $r=r_\ast^s$ \\
\hline
$\M C^u$ & $ \vert+\bullet\hspace{-4pt}\bullet\hspace{2pt}+ \rangle$ & $E\geq1$ & $l=l^u$ & $r=r_\ast^u$ \\
\hline
$\M C_\isco$ & $\vert+ \bullet\hspace{-4pt}\bullet\hspace{-4pt}\bullet\hspace{2pt}-\rangle$ & $E= E_{\isco \pm}$ & $l= l_{\isco \pm}$ & $r=r_\isco$\\
\hline \hline
\end{tabular}
\label{table:classification}
\end{table}

In case we would like to specify the sign of angular momentum $l$ and (or) radial direction $s_r\in\{-1,+1\}$ to avoid ambiguity, we will use subscripts ``$\pm$'' to represent prograde/retrograde emitters and use subscripts ``$,i/o$'' to represent ingoing/outgoing particles. Then a specific orbit (or a quantity $O$) is labeled like ``$\text{Orbit}_{\pm,i/o}$'' (or ``$O_{\pm,i/o}$'').
Otherwise, the subscripts ``$\pm$'' and (or) ``$,i/o$'' may be dropped out for simplicity.

\section{Photon escapes from equatorial emitters}\label{sec:photonescapes}
\subsection{Photon escaping probability}
In \cite{Yan:2021ygy}, a pair of local emission angles from an equatorial emitter has been defined, and the critical emission angles for photons that can escape to infinity have been derived. Here we review these angles and then define the PEP and MOB.
We use $k^\m$ and $p^u$ to represent the four-momentum of emitters and photons, respectively, and we use subscript ``$s$'' to denote the conserved quantities for emitters (light source) while quantities without subscript are for photons. In this work, we will only consider emitters with $l_s\leq l_H$ and $E_s> 0$.
We introduce the emitter's local rest frame (LRF) based on the zero-angular-momentum observer (ZAMO) frame. The ZAMO frame $e_{(\m)}$ is given by \cite{Bardeen:1972fi}
\be
\label{LNRFdef}
e_{(0)}=\sqrt{\f{\Xi}{\D\S}}(\partial_0+\frac{2ar}{\Xi}\partial_3),\quad
e_{(1)}=\sqrt{\f{\D}{\S}}\partial_1,\quad
e_{(2)}=\f{1}{\sqrt{\S}}\partial_2,\quad
e_{(3)}=\sqrt{\f{\S}{\Xi\sin^2\th}}\partial_3.
\ee
Then the 3-velocity and the boost factor of an emitter relative to the ZAMO are given by
\be
\label{velocityandboost}
v_s^{(i)}=\frac{k^\mu e_\mu^{(i)}}{k^\mu e_\mu^{(0)}}\Bigg|_{x^i=x^i_s},\quad
(i=1,2,3),\qquad
v_s=\sqrt{(v_s^{(1)})^2+(v_s^{(3)})^2},\qquad
\g_s=\f{1}{\sqrt{1-v_s^2}}.
\ee
Then we define the LRF of the emitter $\s_{[\m]}$ by
\begin{subequations}
\label{lrf}
\bea
\label{lrf0}
\s_{[0]}&=&\g_s[e_{(0)}+v_s^{(1)}e_{(1)}+v_s^{(3)}e_{(3)}]|_{x^i=x^i_s},\\
\label{lrf1}
\s_{[1]}&=&\f{1}{v_s}[v_s^{(3)}e_{(1)}-v_s^{(1)}e_{(3)}]|_{x^i=x^i_s}\\
\label{lrf2}
\s_{[2]}&=&e_{(2)}|_{x^i=x^i_s},\\
\label{lrf3}
\s_{[3]}&=&\g_s[v_s e_{(0)}+\f{1}{v_s}(v_s^{(1)}e_{(1)}+v_s^{(3)}e_{(3)})]|_{x^i=x^i_s}.
\eea
\end{subequations}
Then the local emission angles $(\a,\b)$ is defined by
\be
\label{localangles}
\a\equiv\arccos\br{\f{p_s^{[3]}}{p_s^{[0]}}}\in[0,\pi],\qquad
\b\equiv\arcsin\br{\f{p_s^{[1]}}{\sqrt{(p_s^{[1]})^2+(p_s^{[2]})^2}}}\in[-\f{\pi}{2},
\f{\pi}{2}],
\ee
where $p^{[a]}_s=p^\mu\s_\mu^{[a]}|_{x^i=x_s^i}$.

\begin{figure}[h]
\centering
\includegraphics[width=6cm]{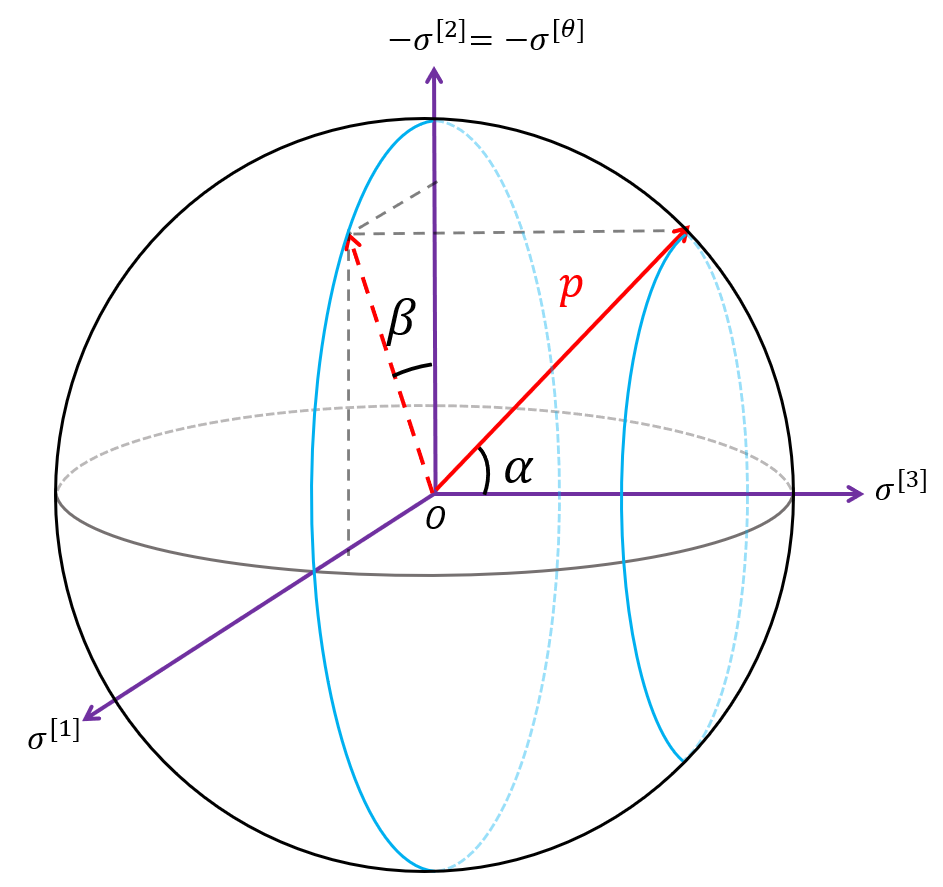}
\caption{Local emission angles ($\a$, $\b$) on an emitter's sky \cite{Igata:2021njn}.}
\label{fig:emittersky}
\end{figure}

Critical photon emissions correspond to unstable double roots of the null radial potential, $\M R(\td r) = \M R^{\prime}(\td r) = 0$. The solutions are given by \cite{Gralla:2019drh}
\bea
\label{criticalpara}
\td{\l}(\td r) = a+\frac{\td r}{a}\left[\td r-\frac{2\td \Delta(\td r)}{\td r-1}\right],\quad
\td{\eta}(\td r) = \frac{\td r^3}{a^2}\left[\frac{4\td\Delta (\td r)}{(\td r-1)^2}-\td r^2\right],
\eea
and the double roots $\td r$ are in the range of $r_{p+}<\td r<r_{p-}$. Hereafter, the quantities with tildes are for those of the critical photon emissions. Critical emission angles $(\td \alpha,\td\b)$ are obtained by plugging \eqref{criticalpara} into \eqref{localangles}.

We assume that the emitter emits monochromatic photons isotropically in its LRF \eqref{lrf}. Some photons are captured by the black hole, and others escape to infinity. The boundary of the escaping and captured regions is called the critical curve on the emitter's sky, lined up by the critical emission angles $(\td \a, \td \b)$. The photon captured region contains the ``direction to the black hole center" \cite{Gates:2020els}, $p^\m_\bullet$, which corresponds to ingoing photons with $\l=\eta=0$.
Let $\mathcal{A}_{e}$ and $\mathcal{A}_{c}$ respectively be the areas of the photon escaping and captured regions on the emitter's sky of unit radius.
It is convenient to compute the area of the interior region of the critical curve $\M A_{in}$, which equals to $\M A_e/\M A_c$ when $p^\m_\bullet$ is outside/inside the critical curve, respectively.
The interior area $\M A_{in}$ can be computed by \cite{Gates:2020els,Igata:2021njn}
\be
\label{AreaCapture}
\mathcal{A}_{in}=\int_{in}\td \rho d\td\rho d\td\varphi
=\int_{in}\frac{1}{2}\td\rho^2d\td\varphi,
\ee
where
\be
\label{plannercoord}
\rho=\sqrt{2(-\cos\a+1)},\qquad
\varphi=\frac{\pi}{2}+\b.
\ee
Then we define the PEP by \cite{Ogasawara:2019mir,Gates:2020els}
\be
\label{defEP}
P\equiv\frac{\mathcal{A}_e}{4\pi}=\frac{1-\mathcal{A}_c}{4\pi}.
\ee

\subsection{Maximum observable blueshift}\label{sec:mob}
For a photon with energy $E$ reaching asymptotic infinity, the redshift factor $g$ and blueshift factor $z$ are defined by
\be
\label{defredshift}
g\equiv \f{E}{p_s^{[0]}},\qquad
z\equiv 1-\f{1}{g}.
\ee
Using Eqs.~\eqref{kerrPr}--\eqref{kerrThpotential} and Eq.~\eqref{lrf} and letting $\s_r=\pm_r$ in Eq.~\eqref{kerrPr} for photon motions, we get
\be
\label{blueshift}
z(\l,\eta)=1-\g_s\frac{\sqrt{r_s\xi_s^3\D(r_s)}-\s_rv_s^{(r)}\xi_s\sqrt{\D(r_s)\M R_p(r_s)}-\br{v_s^{(\phi)}\sqrt{r_s^3\xi_s}\D(r_s)+2a\sqrt{r_s\xi_s\D(r_s)}}\l}
{r_s\xi_s\D(r_s)},
\ee
where $v_s^{(r)}$, $v_s^{(\phi)}$ and $\g_s$ are defined in \eqref{velocityandboost}, and
\bea
\label{xis}
&&\xi_s=r_s^3+a^2(r_s+2),\\
\label{photonpoten}
&&\M R_p(r_s)=\f{\M R(r_s)}{E^2}=(r_s^2+a^2-a\l)^2-\D(r_s)[\eta+(\l-a)^2].
\eea
Then photons with $z>0$ have net blueshift at infinity. The MOB $z_\mob$ is defined by the maximum value of the blueshifts among all the escaping photons emitted at a given position along the emitter's orbit.
Expressing $v^{(r)}$, $v^{(\phi)}$ and $\g_s$ in terms of the emitters' parameters, 
then we have
\be
z(\l,\eta)=1-\frac{E_s\sqrt{\chi_s^2}}{r_s\D(r_s)}+\text{sign}(\chi_s)
\frac{r_s[(r_s-2)l_s+2aE_s]\l+\s_r s_r\sqrt{\M R_s(r_s)\M R_p(r_s)}}{r_s^2\D(r_s)},
\ee
where
\bea
&&\chi_s=r_s^3+a^2(r_s+2)-2a\frac{l_s}{E_s},\\
&&\M{R}_s(r_s)=[E_s(r_s^2+a^2)-al_s]^2-\D(r_s)[(l_s-aE_s)^2+r_s^2].
\eea
In this work, we only consider emitters with $l_s\leq l_H$ and $E_s>0$, then for $r_s\geq r_H$ we have [see from Eqs.~\eqref{horizon} and \eqref{thermbound}]
\be
\chi_s\geq r_H^3+a^2(r_H+2)-2a\frac{l_H}{E_H}=0
\ee
with the equality being obtained at the horizon, $r_s=r_H$.

To find the maximum value of $z(\l,\eta)$, we first compute the partial derivative of $z$ over $\eta$. The result is
\be
\label{pzpeta}
\frac{\partial z}{\partial\eta}
=-\s_r s_r\text{sign}(\chi_s)\frac{1}{2r_s^2}\sqrt{\frac{\M R_s(r_s)}{\M R_p(r_s)}}.
\ee
At the horizon $r_s=r_H$, we have $\p z/\p\eta=0$. Outside the horizon $r_s>r_H$, the sign of $\p z/\p\eta$ depends on the radial motion directions of the photon and emitter, $\p z/\p\eta\propto-\s_r s_r$. When $\s_r s_r=\pm1$, the partial derivative $\p z/\p\eta\lessgtr0$, thus the blueshift $z$ decreases/increases monotonically with $\eta$. Therefore, the MOB $z_{\mob}$ would be obtained at a certain point $(\l,\eta)$ residing at the lower or upper bounds of $\eta(\lambda)$ when $\s_r s_r=1$ or $-1$, respectively.

\begin{figure}[!thbp]
\includegraphics[scale=0.3]{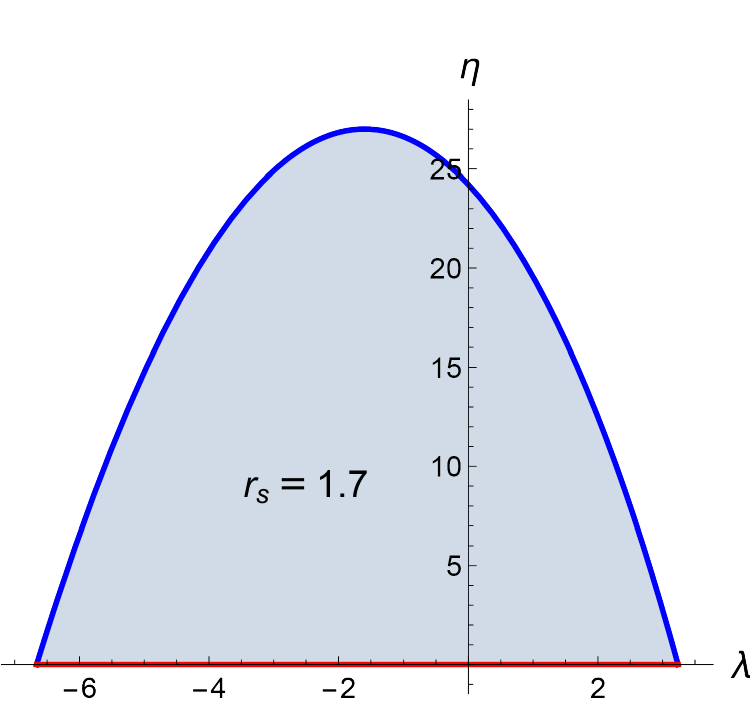}
\includegraphics[scale=0.3]{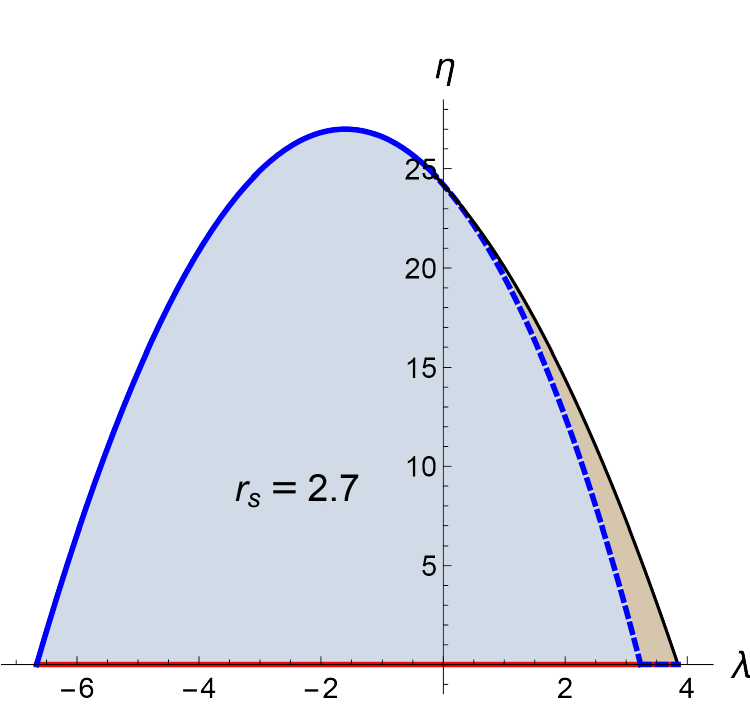}
\includegraphics[scale=0.3]{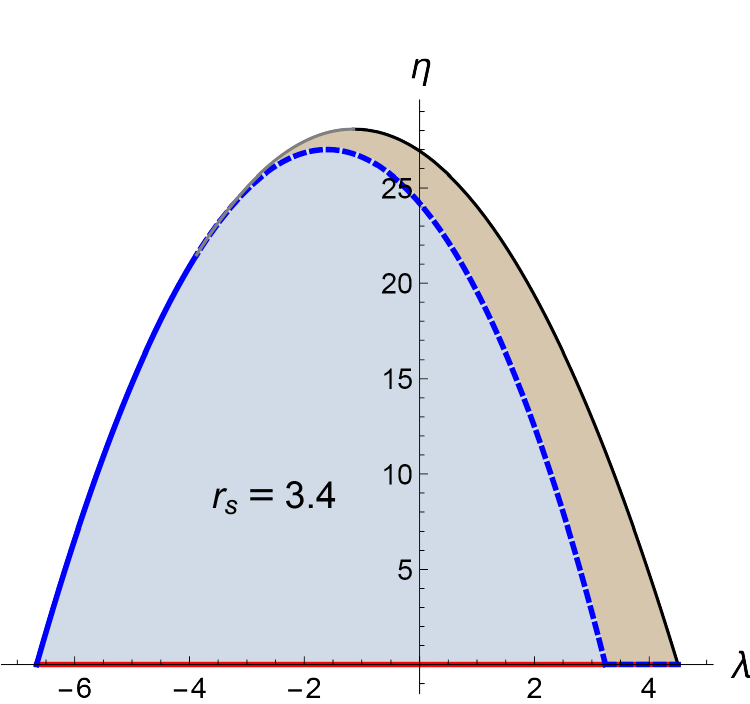}
\includegraphics[scale=0.3]{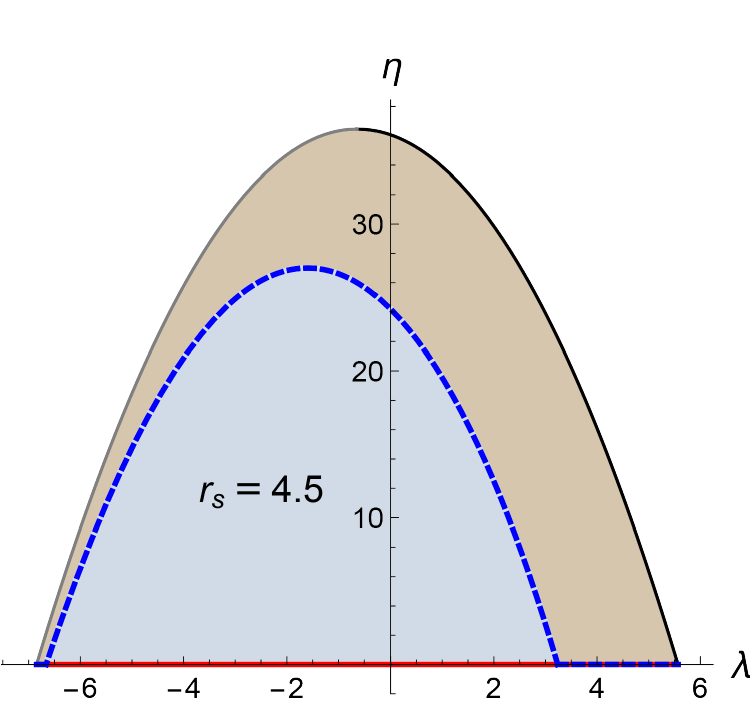}
\put(5,25){\includegraphics[scale=0.6]{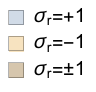}}
\caption{An example of the photon escaping regions in the $\eta-\lambda$ plane for black hole spin $a=0.8$, where the emitter radii respectively belong to the four separate cases in Table \ref{table:escapeparameter}.
The boundary between orange region ($\s_r=\pm1$) and the blue region $(\s_r=+1)$ is described by the function $\tilde\l(\td\eta)$, the gray and black boundary lines denote $\l_2(r_s;\eta)$ and $\l_1(r_s;\eta)$, respectively.
The blue and red curves (solid has $\s_r=+1$ and dashed has $\s_r=-1$) denote the bounds of $\eta(\l)$ where the MOB is obtained for ingoing $(s_r=-1)$ and outgoing $(s_r=+1)$ emitters, respectively.
}
\label{fig:photonregion}
\end{figure}

Next, we analyze the blueshift $z(\l,\eta)$ in the impact parameter region of escaping photons and find out the maximum value of $z(\l,\eta)$ for each emitter by numerically run over the corresponding parameter bound $[\l,\eta(\lambda),\s_r]$. The photon escaping parameter region has been clarified in \cite{Igata:2021njn}. Here we show their results (Table I in \cite{Igata:2021njn}) with our notations in Table \ref{table:escapeparameter},
where
\be
\eta_{\text{max}}=\f{r_s^3}{r_s-2},
\ee
and [eliminating $\td r$ from Eq.~\eqref{criticalpara}]
\be
\td\l_{1,2}=\td\l_{1,2}(\td\eta)=\td\l[\td r_{1,2}(\td\eta)], \quad (\td r_2\geq \td r_1)
\ee
and [solving $\M R_p(r)=0$]
\be
\label{lambda12}
\l_{1,2}(r;\eta)=\f{-2ar\pm\sqrt{r\D(r)[r^3-\eta(r-2)]}}{r(r-2)}.
\ee
with the subscripts ``1, 2'' corresponding to the plus and minus signs, respectively.
In Fig.~\ref{fig:photonregion}, we show an example of the photon escaping regions in the $\eta-\lambda$ plane. We find that, the MOB for outgoing emitters, $z_{\mob,o}$, is obtained at the bound $(\eta,\s_r)=(0,+1)$ which are denoted with solid red lines in Fig.~\ref{fig:photonregion}. While the MOB for ingoing emitters, $z_{\mob,i}$, is obtained at the bounds $(\eta, \s_r)=[\td\eta(\td \l),+1]$, $(\eta, \s_r)=[\td\eta(\td \l),-1]$ or $(\eta,\s_r)=(0,+1)$, which are denoted with solid ($\s_r=+1$) and dashed ($\s_r=-1$) blue lines in Fig.~\ref{fig:photonregion}.

Now we explain the corrections in $z_{\mob,i}$ of Refs.~\cite{Igata:2021njn} and \cite{Yan:2021ygy}. In \cite{Igata:2021njn} and \cite{Yan:2021ygy}, the MOB of ingoing and outgoing emitters ($s_r=\pm1$) was all obtained for outward escaping photons with $\s_r=+1$. However, from Eq.~\ref{pzpeta}, Table \ref{table:escapeparameter} and Fig.~\ref{fig:photonregion}, we find that for ingoing emitters with $r_s>r_p$ the MOB is obtained instead for inward escaping photons with $\s_r=-1$.

\begin{table}[h]
\centering
\caption{Parameter region for all escaping photons from a source at $r=r_s$ \cite{Igata:2021njn}.}
\begin{tabular}{c| c| c| c }
\hline\hline
Case& $\eta$ & $\lambda$ ($\s_r=+1$) & $\lambda$ ($\s_r=-1$)\\
\hline
$r_H<r_s<r_{p+}$ & $0\leq\eta\leq27$ & $\td\l_2<\lambda<\td\lambda_1$ & -\\
\hline
$r_{p+}<r_s<3$ & $0\leq\eta<\td\eta(r_s)$ & $\td\l_2<\l\leq\l_1(r_s;\eta)$ &
$\td\l_1<\l<\l_1(r_s;\eta)$\\
& $\td\eta(r_s)\leq\eta\leq27$ & $\td\l_2<\l<\td\l_1$ & -\\
\hline
$3\leq r_s< r_{p-}$ & $0\leq\eta<\td\eta(r_s)$ & $\td\l_2<\l\leq \l_1(r_s;\eta)$ & $\td\l_1<\l\leq \l_1(r_s;\eta)$ \\
& $\td\eta(r_s)\leq\eta<27$ & $\l_2(r_s;\eta)\leq\l\leq\l_1(r_s;\eta)$ &
$\l_2(r_s;\eta)<\l<\td\l_2$, $\td\l_1<\l<\l_1(r_s;\eta)$\\
& $27\leq\eta\leq\eta_\text{max}$ & $\l_2(r_s;\eta)\leq\l\leq\l_1(r_s;\eta)$& $\l_2(r_s;\eta)\leq\l\leq\l_1(r_s;\eta)$\\
\hline
$r_s\geq r_{p-}$ & $\td\eta(r_s)\leq\eta<27$ & $\l_2(r_s;\eta)\leq\l\leq\l_1(r_s;\eta)$ &
$\l_2(r_s;\eta)<\l<\td\l_2$, $\td\l_1<\l<\l_1(r_s;\eta)$\\
& $27\leq\eta\leq\eta_\text{max}$ & $\l_2(r_s;\eta)\leq\l\leq\l_1(r_s;\eta)$& $\l_2(r_s;\eta)\leq\l\leq\l_1(r_s;\eta)$\\
\hline\hline
\end{tabular}
\label{table:escapeparameter}
\end{table}

\section{PEP and MOB for emitters on different orbits}\label{sec:PEPandMOB}
Now we study the PEP and MOB of photon emissions from emitters on different orbits, which depend on the black hole spin $a$, and the emitters' parameters $(r_s, E_s, l_s, s_r,)$.
Following \cite{Igata:2021njn}, we introduce $P\geq1/2$ and $z_\mob\geq0$ as indicators of an emitter's observability. We use $r_{0}$ to denote the radii at which $P=1/2$ and use $r_{z}$ to represent the radii where $z_\mob=0$.
The results are shown in Figs.~\ref{fig:mtisco}--\ref{fig:ptbd}. Each pair of PEP and MOB curves can reflect the variation in the brightness of an emitter along its orbit.
We can see that the general feature of the results for prograde ($l_s>0$) and retrograde ($l_s<0$) emitters are very similar. However, as the black hole spin $a$ varies, the changing trends of the results for prograde and retrograde emitters are different (see Figs.~\ref{fig:mtisco} and \ref{fig:mp}). We also find that for outgoing ($s_r=+1$) emitters, the PEP is larger than one half and the MOB is positive, that is $P_{,o}>1/2$ and $z_{\mob ,o}>0$. This means that all outgoing emitters are well observable.
Therefore, in the following, we will pay more attention to the prograde and ingoing emitters, which have $l_s>0$ and $s_r=-1$.

\subsection{Emitters on marginal trapped orbits from the ISCO}
\begin{table}[h]
\centering
\caption{Numerical values of some characteristic radii and the parameter $k_\pm$ for ingoing marginal trapped orbits from the ISCO ($\M {MT}_\isco$).}
\begin{tabular}{c|c|c|c|c|c|c|c|c|c}
\hline\hline
$a$ & 0 & 0.1 & 0.3 & 0.5 & 0.7 & 0.9 & 0.99 & 0.999 & 0.9999 \\
\hline
$r_H$ & 2.000 & 1.995 & 1.954 & 1.866 & 1.714 & 1.436 & 1.141 & 1.045 & 1.014 \\
\hline
$r_{\isco+}$ & 6.000 & 5.669 & 4.979 & 4.233 & 3.393 & 2.321 & 1.454 & 1.182 & 1.079 \\
\hline
$r_{0+}$ & 3.464 & 3.343 & 3.074 & 2.756 & 2.365 & 1.801 & 1.281 & 1.106 & 1.039 \\
\hline
$r_{z+}$ & 2.883 & 2.775 & 2.546 & 2.285 & 1.974 & 1.544 & 1.167 & 1.053 & 1.017 \\
\hline
$r_{s+}^\prime$ & 3.528 & 3.374 & 3.042 & 2.674 & 2.478 & 1.681 & 1.207 & 1.066 & 1.022\\
\hline
$r_{p+}$ &3.000 & 2.882 & 2.630 & 2.347 & 2.013 & 1.558 & 1.168 & 1.052 & 1.016\\
\hline
$k_+$ & 0.433 & 0.436 & 0.443 & 0.452 & 0.463 & 0.479 & 0.494 & 0.497 & 0.496 \\
\hline
$r_{\isco-}$ & 6.000 & 6.323 & 6.949 & 7.555 & 8.143 & 8.717 & 8.972 & 8.997 & 9.000 \\
\hline
$r_{0-}$ & 3.464 & 3.580 & 3.788 & 3.975 & 4.140 & 4.282 & 4.336 & 4.337 & 4.340 \\
\hline
$r_{z-}$ & 2.883 & 2.986 & 3.180 & 3.362 & 3.535 & 3.699 & 3.771 & 3.778 & 3.778\\
\hline
$r_{s-}^\prime$& 3.528 & 3.681 & 3.972 & 4.247 & 4.513 & 4.769 & 4.881 & 4.892 & 4.892\\
\hline
$r_{p-}$ & 3.000 & 3.113 & 3.329 & 3.532 & 3.725 & 3.910 & 3.991 & 3.999 & 3.999\\
\hline
$k_-$ & 0.433 & 0.430 & 0.425 & 0.422 & 0.420 & 0.422 & 0.429 & 0.432 & 0.433 \\
\hline\hline
\end{tabular}
\label{table:datas}
\end{table}
\begin{figure}[ht]
\includegraphics[scale=0.48]{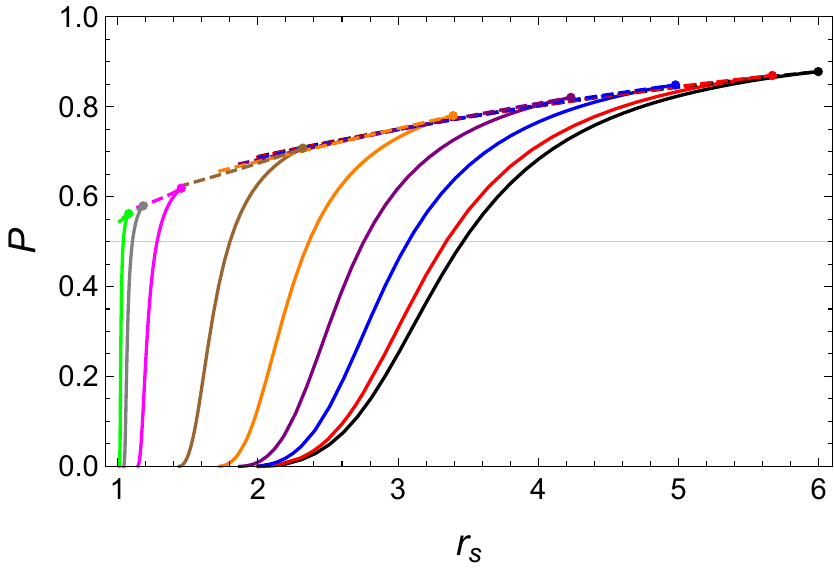}\,\,
\includegraphics[scale=0.48]{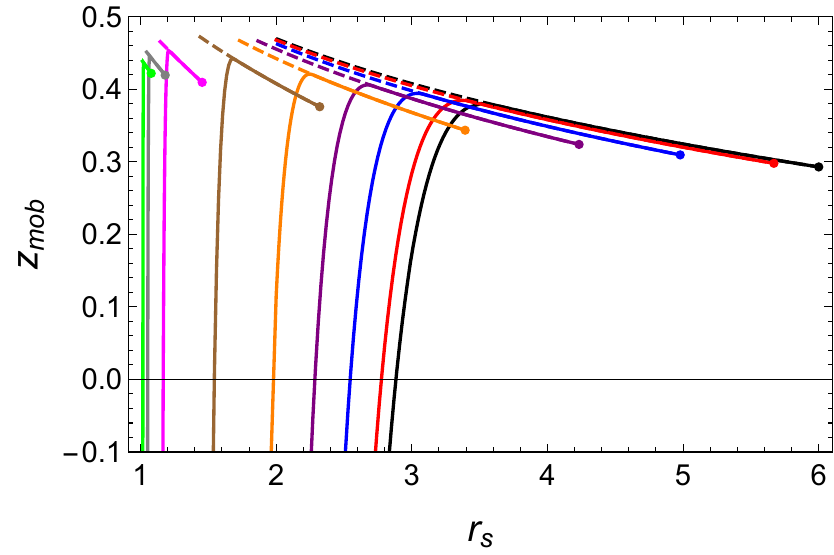}\,\,\\
\includegraphics[scale=0.48]{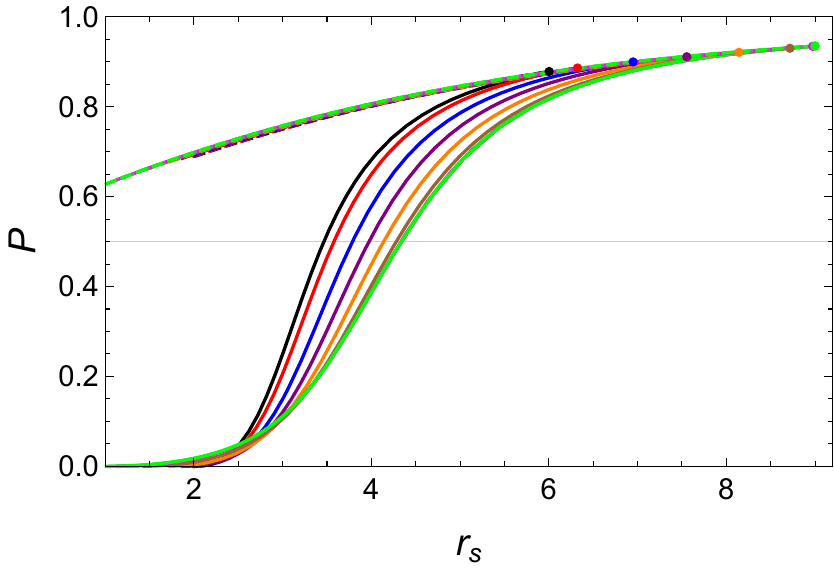}\,\,
\includegraphics[scale=0.48]{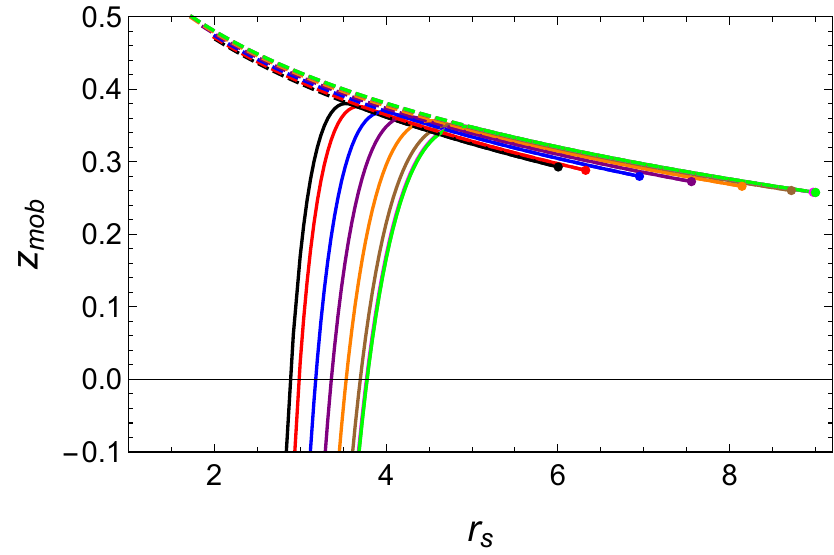}\,\,
\put(5,25){\includegraphics[scale=0.75]{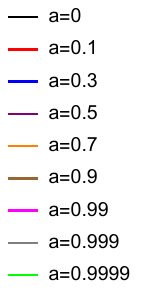}}
\caption{PEP $P(r_s)$ and MOB $z_\mob(r_s)$ of $\M{MT}_\isco$ emitters for several values of $a$. The above row is for prograde ($l_s>0$) emitters, while the below row is for retrograde ($l_s<0$) emitters. Solid curves are for ingoing ($s_r=-1$) emitters, while dashed curves are for outgoing ($s_r=+1$) emitters. The dots are for the results at the ISCO $r_\isco$.}
\label{fig:mtisco}
\end{figure}
\begin{figure}[!hb]
\includegraphics[scale=0.48]{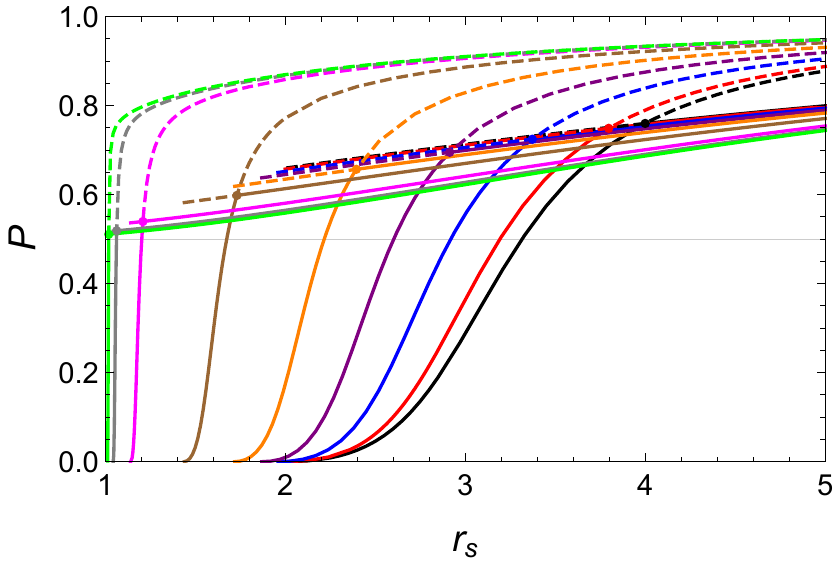}\,\,
\includegraphics[scale=0.48]{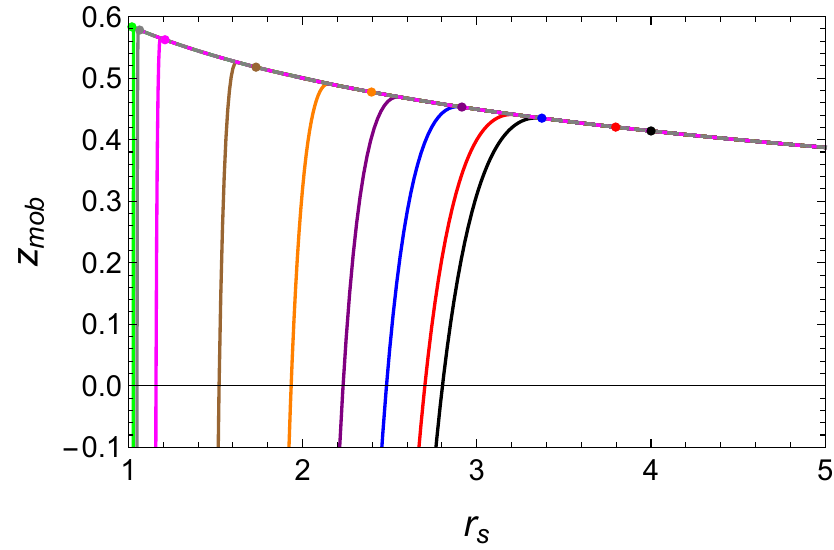}\,\,\\
\includegraphics[scale=0.48]{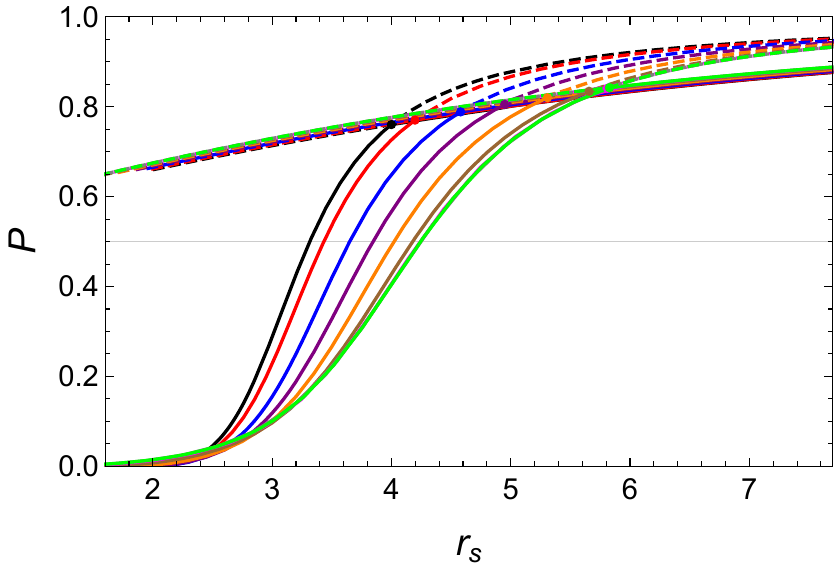}\,\,
\includegraphics[scale=0.48]{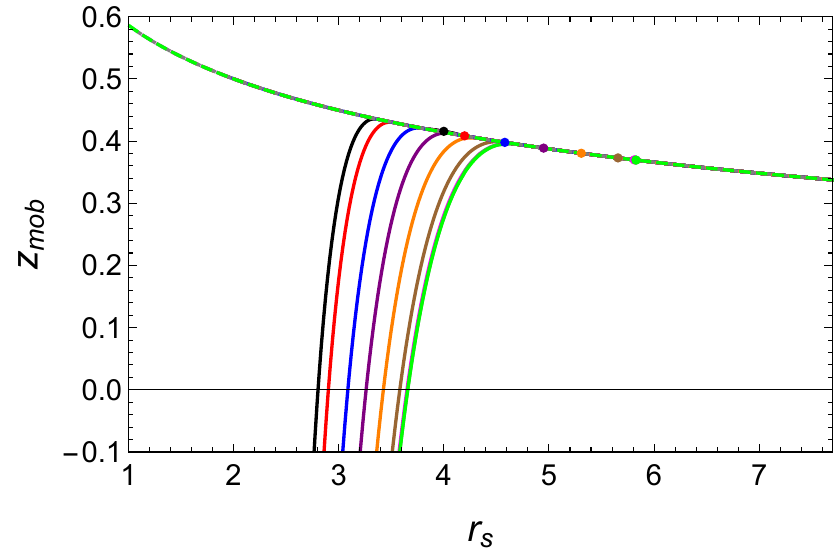}\,\,
\put(5,25){\includegraphics[scale=0.75]{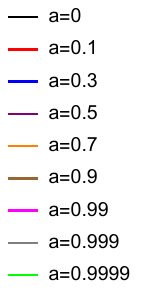}}
\caption{PEP $P(r_s)$ and MOB $z_\mob(r_s)$ for (anti-)$\M{MP}$ emitters with $E_s=1$ for several values of $a$. The above row is for prograde ($l_s>0$) emitters, while the right below row is for retrograde ($l_s<0$) emitters. Solid curves are for ingoing ($s_r=-1$) emitters, while dashed curves are for outgoing ($s_r=+1$) emitters. The dots are for the results at the unstable double root $r_\ast$. }
\label{fig:mp}
\end{figure}

\begin{figure}[h]
\includegraphics[scale=0.48]{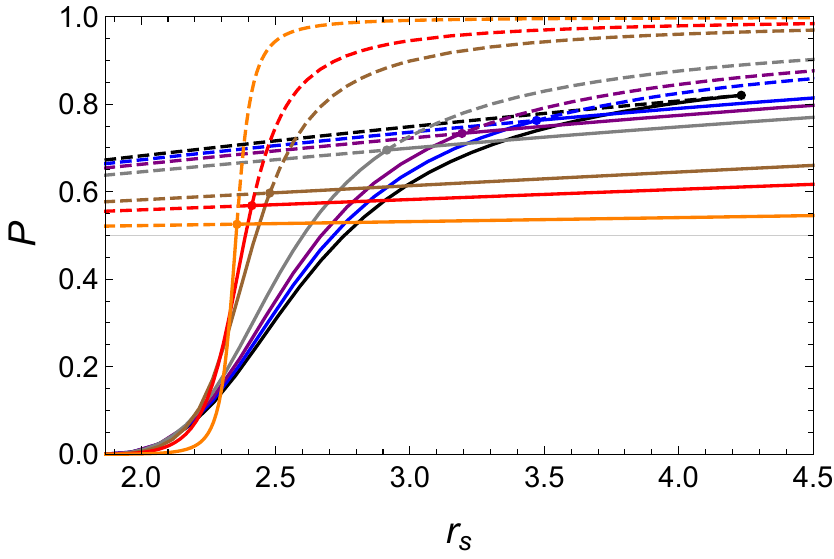}\,\,
\includegraphics[scale=0.48]{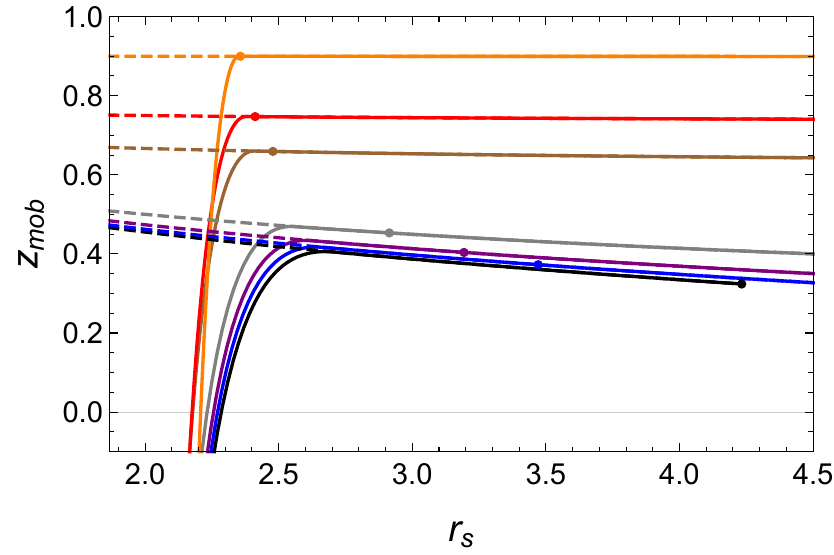}\,\,
\put(5,25){\includegraphics[scale=0.75]{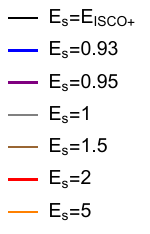}}\\
\includegraphics[scale=0.48]{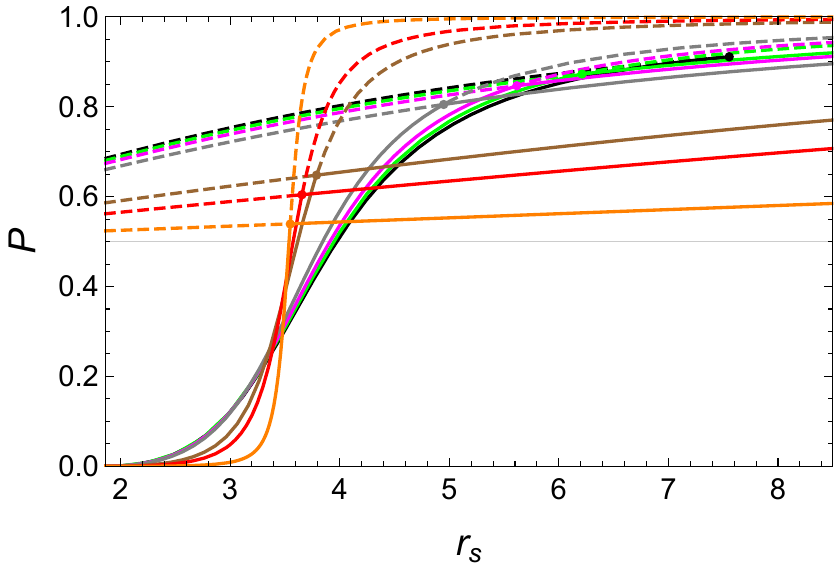}\,\,
\includegraphics[scale=0.48]{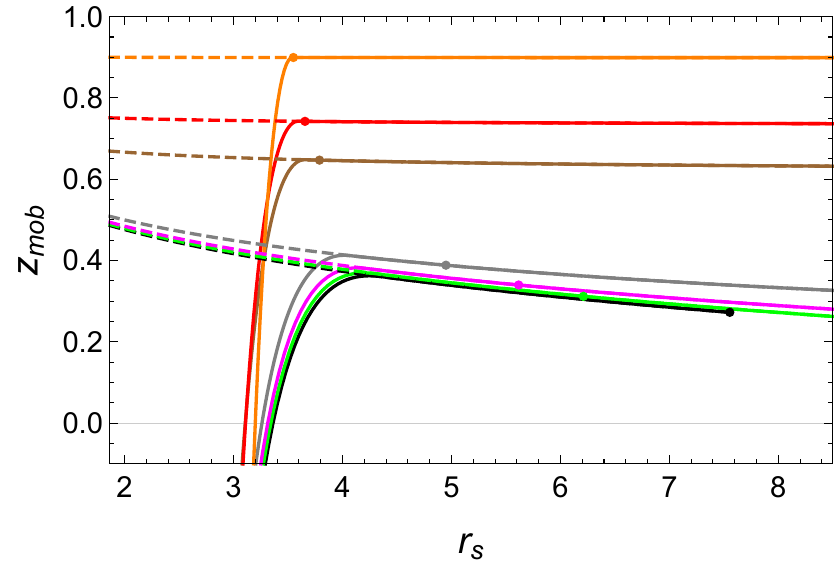}\,\,
\put(5,25){\includegraphics[scale=0.75]{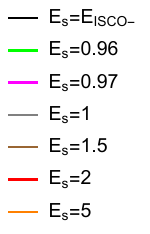}}
\caption{PEP $P(r_s)$ and MOB $z_\mob(r_s)$ for (anti-)$\M{MP}$ and $\M {MT}$ emitters with various $E_s$ for $a=0.5$, where $E_{\isco+}=0.918$ and $E_{\isco-}=0.955$. The above row is for prograde ($l_s>0$) emitters, while the below row is for retrograde ($l_s<0$) emitters. Solid curves are for ingoing ($s_r=-1$) emitters, while dashed curves are for outgoing ($s_r=+1$) emitters. The dots are for the results at the unstable double root $r_\ast$. Note that the black curves are for $\M{MT}_\isco$, which are displayed for reference.}
\label{fig:mptTogether}
\end{figure}

First we consider the marginal trapped emitters from the ISCO $\M {MT}_\isco$, which have $E_s=E_{\isco}$ and $l_s=l_{\isco}$. The PEP $P(r_s)$ and MOB $z_\mob(r_s)$ only depend on black hole spin $a$.
For several values of $a$, the results of $P(r_s)$ and $z_\mob(r_s)$ are shown in Fig.~\ref{fig:mtisco}.
As the source radius $r_s$ is decreased from the ISCO for each given spin, the PEP $P_{,i}(r_s)$ along ingoing orbits decrease rapidly (solid curve in Fig.~\ref{fig:mtisco}) while the PEP $P_{,o}(r_s)$ along outgoing orbits decrease much gently (dashed curve in Fig.~\ref{fig:mtisco}). In addition, as $r_s$ is decreased from the ISCO, the MOB $z_{\mob,o}(r_s)$ increases monotonically while $z_{\mob,i}(r_s)$ increases at the beginning and decreases rapidly after reaching its maximum value. The maximum value of $z_{\mob,i}(r_s)$ is obtained at the radius $r^\prime_s\in(r_p,r_\ast)$ and we can see that $z_{\mob,i}(r_s)=z_{\mob,o}(r_s)$ in the region of $r_s>r_s^\prime$.
To compare the value of $r_{0}$ with the radius of horizon $r_H$ and the radius of the ISCO $r_{\isco}$ for ingoing $\M{MT}_{\isco\pm,i}$ emitters, we define
\be
k_\pm=\f{r_{0\pm}}{r_H+r_{\isco\pm}}.
\ee
We show some numerical results for $k_\pm$ and several characteristic radii in Table~\ref{table:datas}.
We can see that for both prograde and retrograde $\M{MT}_{\isco\pm,i}$ emitters, the values $k_\pm$ are all in the range $0.43\sim0.50$.

The photon escaping from prograde marginal trapped emitters from the ISCO has been studied in \cite{Igata:2021njn}, where the authors found that $r_{z+}<r_{0+}$ and $r_{0+}$ is at roughly the middle point between the horizon and the ISCO. These emitters are called the ``plunging'' emitter from the ISCO in \cite{Igata:2021njn}, which in our notation are specified as $\M{MT}_{\isco+,i}$.
From Fig.~\ref{fig:mtisco} and Table~\ref{table:datas},
we can see that our results for $\M{MT}_{\isco+,i}$ agree with those in \cite{Igata:2021njn} up to a correction\footnote{The corresponding photon parameters for $z_{\mob,i}$ in the region $r_{p+}<r_s<r_{\isco+}$ should be $(\l,\eta,\s_r)=(\l^\prime,0,-1)$ with $\l^\prime$ being a value in the range of $\td\l(r_{p+})<\l^\prime<\l_1(r_s;0)$ [see Eqs.~\eqref{criticalpara} and \eqref{lambda12}], but in \cite{Igata:2021njn} the parameters were taken as $(\l,\eta,\s_r)=(\l_1(r_s,0),0,1)$.} of $z_{\mob,i}$ in the region $r_{p+}<r_s<r_{\isco+}$ (see Sec. \ref{sec:mob} for details).
Moreover, our results also show that for the retrograde ``plunging'' emitters from the ISCO (i.e., $\M{MT}_{\isco-,i}$), the PEP are greater than $1/2$ and the MOB are positive at least until the middle point between the horizon and the ISCO.

\begin{figure}[!thbp]
\includegraphics[scale=0.48]{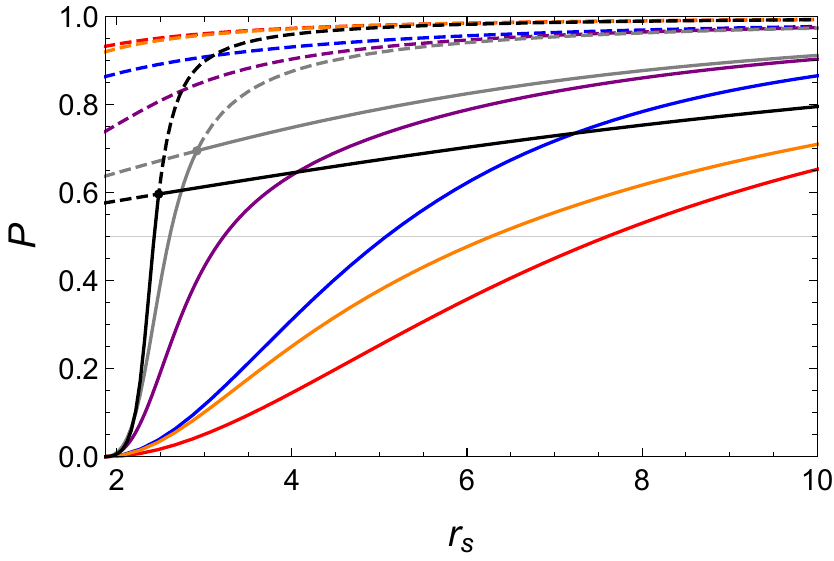}\,\,
\includegraphics[scale=0.48]{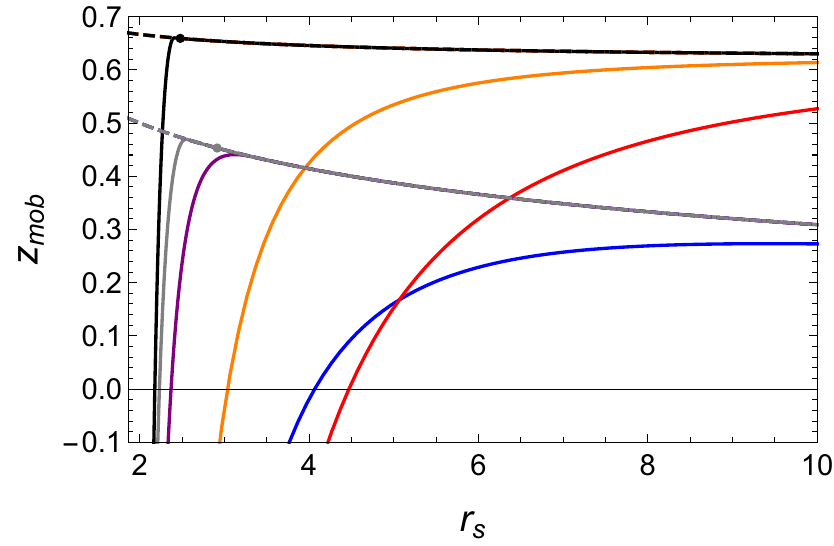}\,\,
\put(5,20){\includegraphics[scale=0.75]{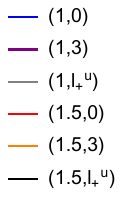}}\\
\begin{overpic}[scale=0.48]{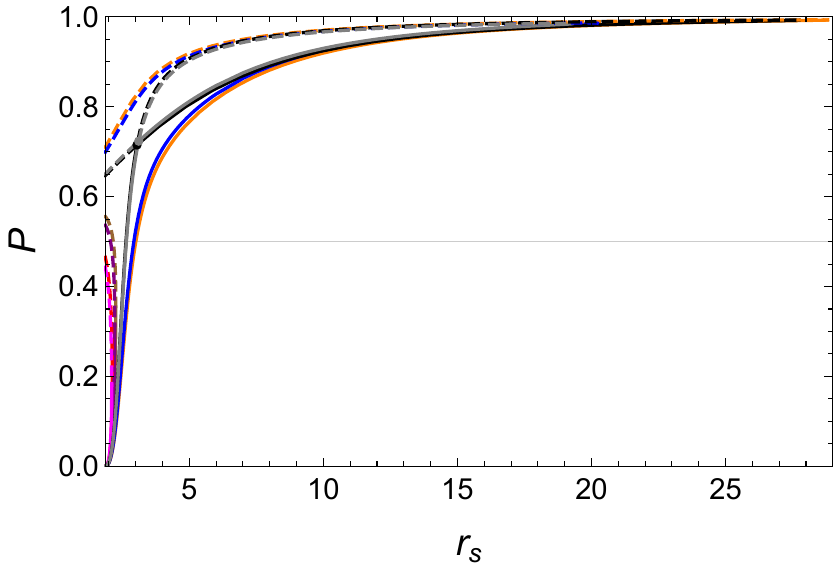}
\put(32,13.75){\includegraphics[scale=0.32]{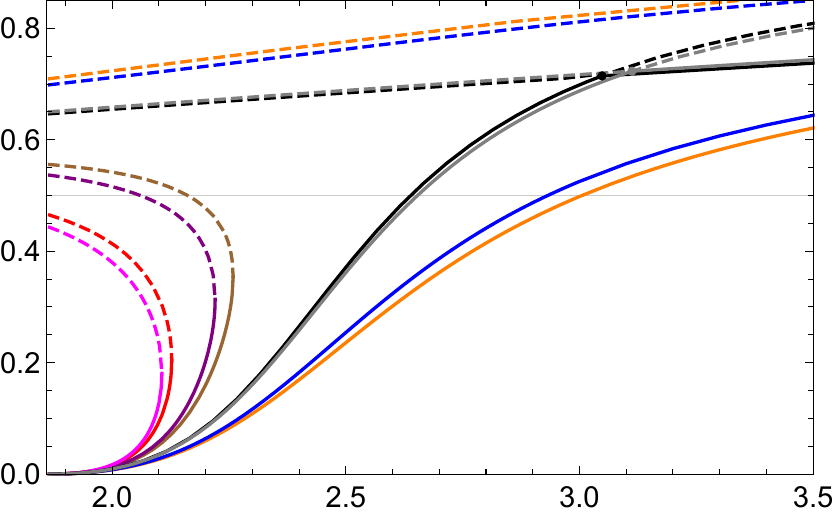}}
\end{overpic}\,\,
\begin{overpic}[scale=0.48]{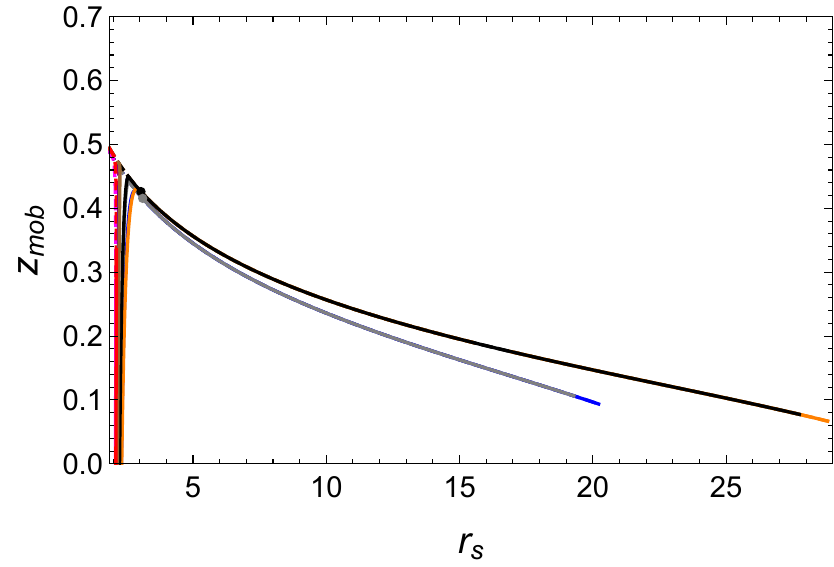}
\put(45,30){\includegraphics[scale=0.26]{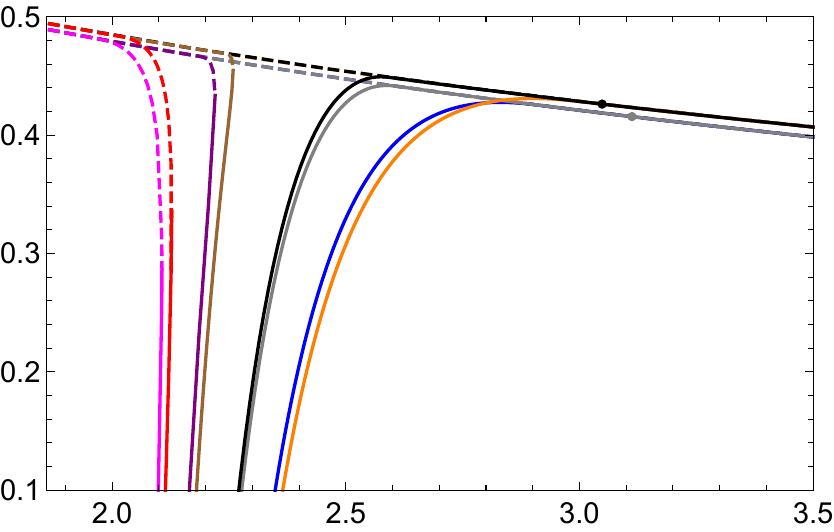}}
\end{overpic}\,\,
\put(5,20){\includegraphics[scale=0.75]{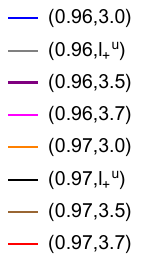}}\\
\includegraphics[scale=0.48]{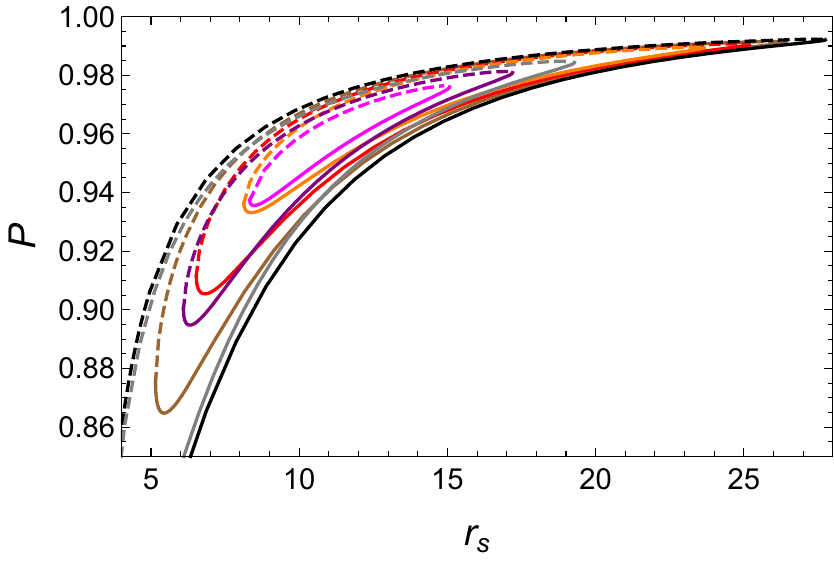}\,\,
\includegraphics[scale=0.48]{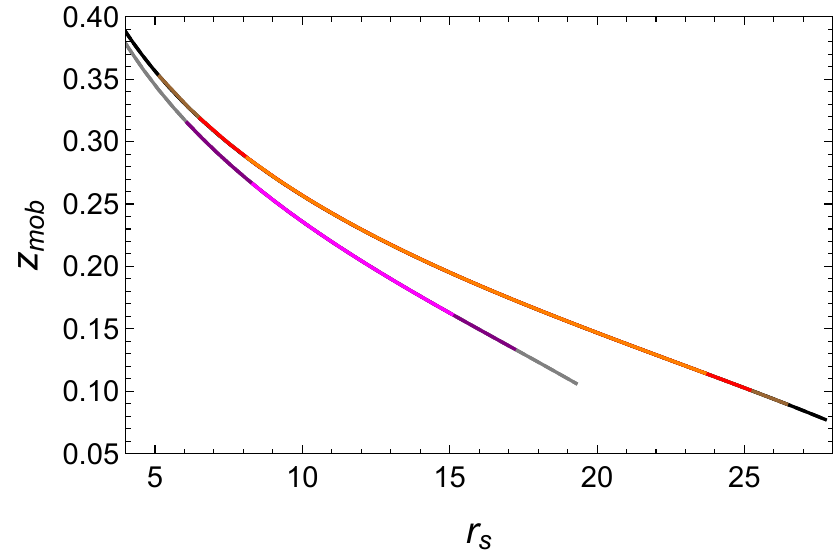}\,\,
\put(5,20){\includegraphics[scale=0.75]{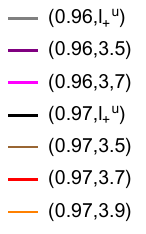}}\\
\includegraphics[scale=0.48]{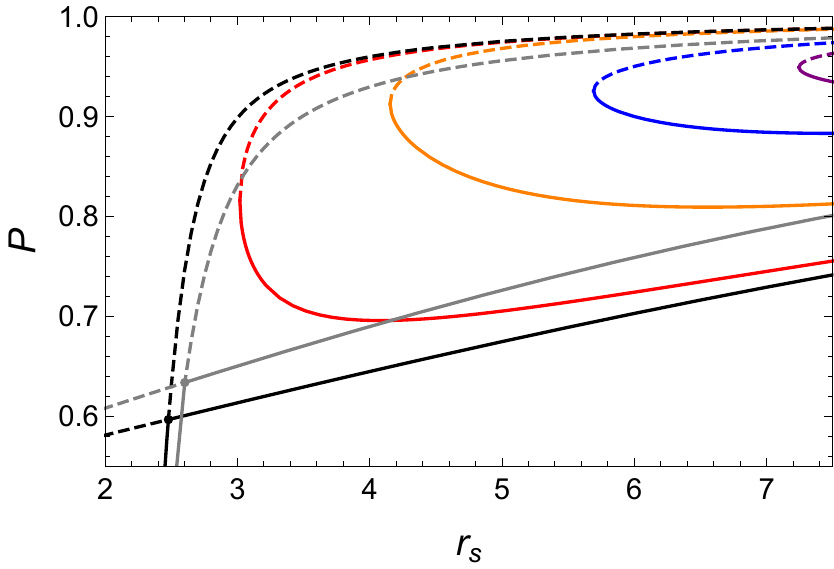}\,\,
\includegraphics[scale=0.49]{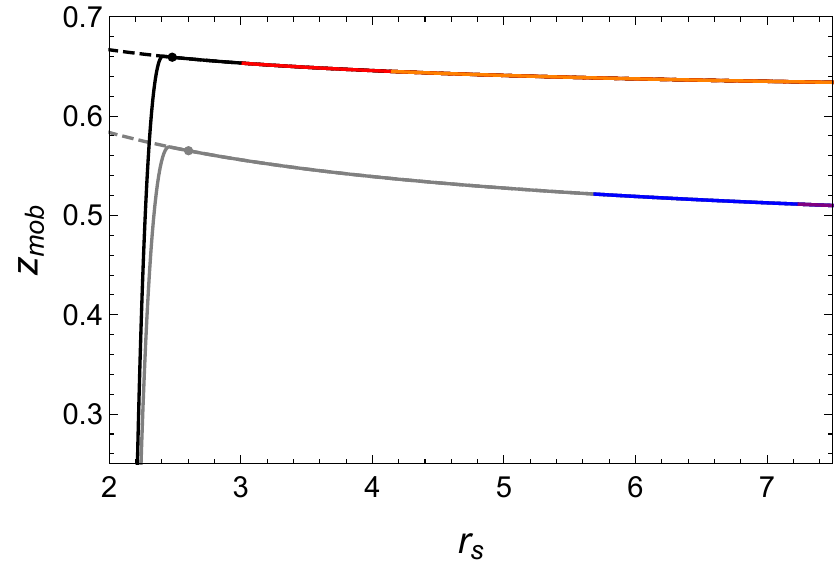}\,\,
\put(5,20){\includegraphics[scale=0.75]{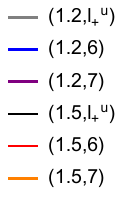}}
\caption{PEP $P(r_s)$ and MOB $z_\mob(r_s)$ for emitters with parameters $(E_s,l_s)$ for $a=0.5$. From top to bottom: the emitters are on prograde (anti-)$\M{P}_+$, $\M {T}_+$, $\M B_+$ and $\M D_+$ orbits, respectively. The relevant numeral values of $l^u_+(E_s)$ are listed in Table~\ref{table:numericalLu}. Solid curves are for ingoing ($s_r=-1$) emitters, while dashed curves are for outgoing ($s_r=+1$) emitters. The dots are for the results at the unstable double root $r_\ast$. Note that the black and gray curves are for marginal emitters displayed for reference.}
\label{fig:ptbd}
\end{figure}

\subsection{Emitters on generic marginal orbits}
Next we consider the generic marginal (anti-)plunging [(anti-)$\M{MP}$] and trapped ($\M{MT}$) emitters which have $l_s=l^u(E_s)$.
Even though the marginal emitters are not physical since a particle can not move across the double root along any of the marginal orbits in a finite proper time,  the orbits of the marginal emitters are the asymptotes for the physically allowed non-marginal emitters, and so do the PEP curves and the MOB curves.
Therefore,  the PEP and the MOB curves for the marginal emitters could serve as key references for diverse types of non-marginal emitters.

The results of $P(r_s)$ and $z_\mob(r_s)$ for marginal emitters depend on the black hole spin $a$ and the emitter energy $E_s$. In Fig.~\ref{fig:mp} we show the results of $P(r_s)$ and $z_\mob(r_s)$ for general marginal emitters with $E_s=1$ for several $a$. Note that, as long as we focus on the near horizon region, the behaviors of PEP and MOB for all marginal [both (anti-)$\M{MP}$ and $\M{MT}$] emitters are similar (see Fig.~\ref{fig:mptTogether} and the second row of Fig.~\ref{fig:ptbd}).
We can see that inside the unstable double root $r_\ast$, the overall feature of $P(r_s)$ and $z_\mob(r_s)$ for general marginal emitters are similar as those for $\M{MT}_\isco$ emitters (see Fig.~\ref{fig:mtisco}).
We also note that the PEP curves for ingoing marginal emitters inside/outside $r_\ast$ connect to their outgoing counter partners outside/inside $r_\ast$ smoothly.

In the following, we focus on ingoing marginal emitters.
We find that $P_{,i}(r_s)$ decreases monotonically as $r_s$ decreases from outside of the double root $r_\ast$ towards the horizon.
In particular, we note that $P_{,i}(r_s)$ decreases slowly when $r_s>r_\ast$, while $P_{,i}(r_s)$ suddenly decreases much rapidly when $r_H<r_s<r_\ast$. In addition, $z_{\mob,i}(r_s)$ increases monotonically as $r_s$ is decreased from outside of $r_\ast$ until reaching $r_s^\prime\in(r_p,r_\ast)$, while as $r_s$ is continued to decrease from $r_s^\prime$ towards horizon $z_{\mob,i}(r_s)$ decreases rapidly.
These features indicate that one could see the image of a marginal ingoing (``plunging'') emitter until it reaching the position of the unstable double root (which is located inside the ISCO). This signature becomes even more striking for emitters on prograde orbits of high-spin black holes.
In Fig.~\ref{fig:mptTogether}, we show the results of $P(r_s)$ and $z_\mob(r_s)$ for both (anti-)$\M{MP}$ and $\M{MT}$ emitters with various $E_s$ for $a=0.5$. 
We find that as $E_s$ is increased from $E_{\isco\pm}$ for the marginal emitters ($\M{MT}$ for $E_\isco <E_s<1$ and $\M{MP}$ for $E_s\geq1$), the PEP and MOB curves move towards smaller radius and the PEP/MOB value at the double root $r_\ast$ decreases/increases. In the near-horizon region $r_H<r_s\lesssim r_\ast$, both $P_{,i}(r_s)$ and $z_{\mob,i}(r_s)$ for emitters with larger $E_s$ decrease faster than those with smaller $E_s$.
Moreover, we always have $P_{,i}(r_\ast)>1/2$ as $E_s$ tends to infinity.
Therefore, we conclude that the above typical observational feature of marginal ``plunging'' emitters is more noticeable for emitters with large $E_s$.
\begin{table}[h]
\centering
\caption{Several numerical values of $[E_s, l^u(E_s)]$ for $a=0.5$.}
\begin{tabular}{c|c|c|c|c|c}
\hline\hline
$E_s$ & 0.96 & 0.97 & 1 & 1.2 & 1.5 \\
\hline
$l^{u}_+$ & 3.186 & 3.245 & 3.414 & 4.411 & 5.767 \\
\hline\hline
\end{tabular}
\label{table:numericalLu}
\end{table}

\subsection{Emitters on non-marginal orbits}
Next we consider the emitters on non-marginal (anti-)plunging (anti-)$\M P$, trapped $\M T$, bounded $\M B$ and deflected $\M D$ orbits. These emitters are divided by the marginal cases in the $(E_s,l_s)$ phase space. As the energy or angular momentum of a non-marginal emitter varies from the one of a marginal emitter, the PEP and MOB curves of the non-marginal emitters deviate from those of the marginal emitter. We study the behaviors of the PEP and MOB of the non-marginal emitters, and compare with the ones of marginal emitters.
The results of $P(r_s)$ and $z_\mob(r_s)$ for prograde emitters with parameters $(E_s,l_s)$ are shown in Fig.~\ref{fig:ptbd}, where we set $a=0.5$.

Now we describe the main feature of these results for the ingoing emitters. We note that for all kinds of emitters at large $r_s$, the PEP and MOB mainly depend on the emitter's energy $E_s$.
For $\M P$ and $\M T$ emitters, as $l_s$ is increased from zero to $l^u$ for each given $E_s$, both the PEP curves $P(r_s)$ and the MOB curves $z_\mob(r_s)$ move towards small radius. Therefore, the $\M{MP}$ (or $\M{MT}$) emitters have the largest observability among all the $\M P$ (or $\M T$) emitters with the same energy $E_s$, and the $\M{MP}$ (or $\M{MT}$) emitters are well observable until they reaching the positions of the unstable double roots.
For $\M B$ emitters, we find that their PEP curves are bounded by those of the marginally trapped emitters\footnote{These may also be treated as the marginal bounded emitters in the range of $r_\ast<r_s<r_t$.} in the range of $r_\ast<r_s<r_t$ with $r_t$ being the radius of the turning point, and so do the MOB curves. Thus we always have $P>1/2$ and $z_\mob>0$ for $\M B$ emitters, which means that the bounded emitters have good observability on their orbits and the observable radial ranges of bounded orbits shrink as $l_s$ is increased from $l^u$ towards $l_H$ (see the second and third rows of Fig.~\ref{fig:ptbd}).
For $\M D$ emitters, we always have $P>1/2$ and $z_\mob>0$ in the whole allowed range $r_s>r_t$. Therefore, the deflected emitters are also well observable along their whole orbits. Note that, as a result of the ``beaming effect", the position of the minima of the PEP curves for the outgoing $\M B$ and $\M D$ emitters deviates from the positions of the turning points.

\subsection{High-spin case}\label{subsec:highspin}
In \cite{Yan:2021ygy}, the photon emissions from the near-horizon ($r\rightarrow r_H$) emitters of a Kerr black hole in the near-extremal limit $a\rightarrow1$ have been studied, in which the PEP and MOB have been computed using the (near-)NHEK metrics. In the near-extremal limit, $a=\sqrt{1-(\e\k)^2}$ with $\e\ll1$, the (near-)NHEK radius $R$ and energy $\hat E$ have been introduced, which are related to the Kerr quantities by
\be
r=1+\e^p R,\qquad
l=2E-\e^p \hat E,\qquad
0<p\leq 1,
\ee
where the NHEK limit has $\k=0$ and $p=2/3$, and near-NHEK limit has $\k=1$ and $p=1$. Note that the ISCO $r_\isco$ is in the NHEK limit and the unstable circular orbits $r_\ast$ are in the near-NHEK limit, and these orbits have the marginal angular momentum $l_s=l^u(E_s)$. On the opposite sides of $l^u(E_s)$ of the $(E_s,l_s)$ phase space, the orbits belong to different classes of motion (see Table~\ref{table:classification}). Next, we will compute the PEP and MOB of photon emissions from the equatorial plane of a high-spin black hole using the Kerr metric.

\begin{figure}[t]
\begin{overpic}[scale=0.48]{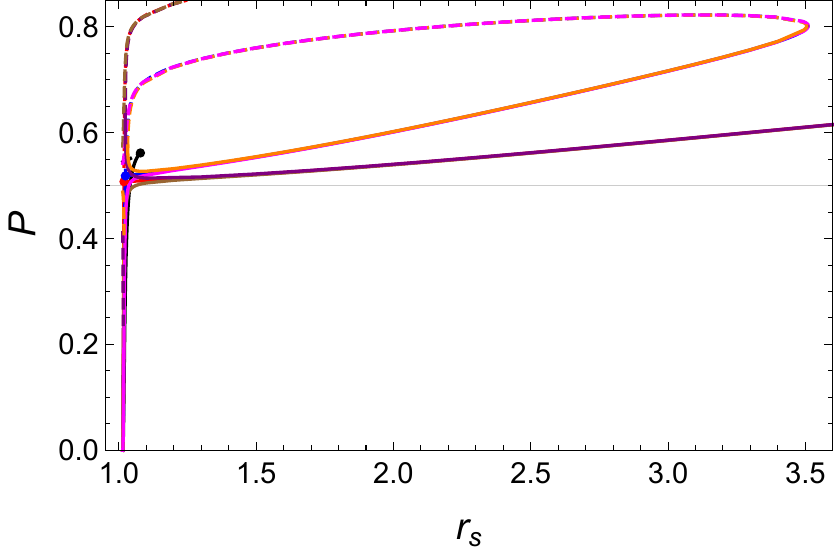}
\put(49,13){\includegraphics[scale=0.24]{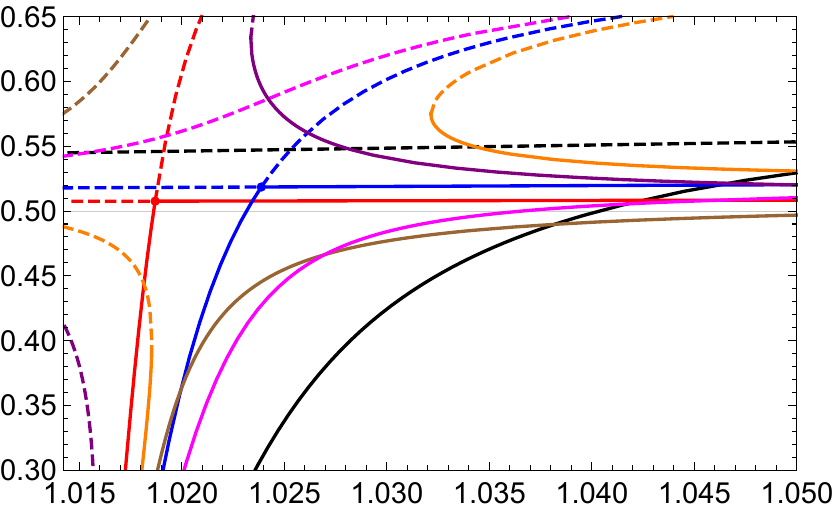}}
\end{overpic}\,
\begin{overpic}[scale=0.48]{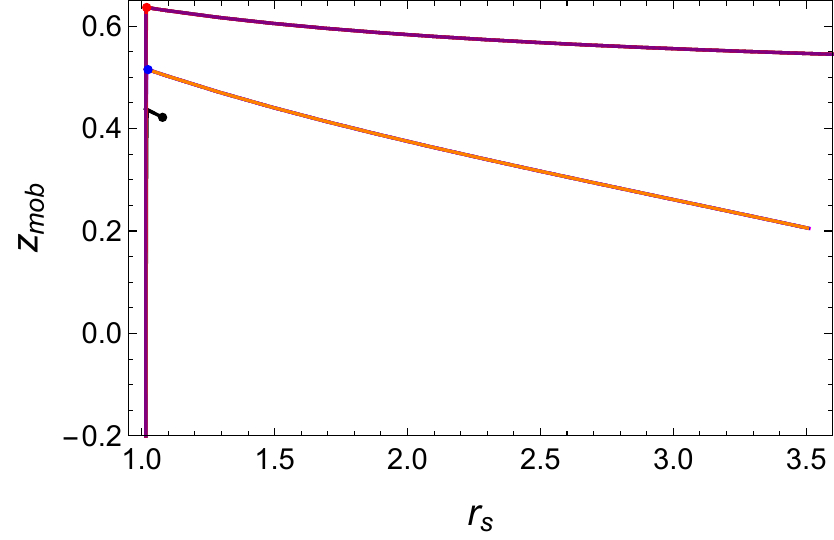}
\put(18.5,13){\includegraphics[scale=0.24]{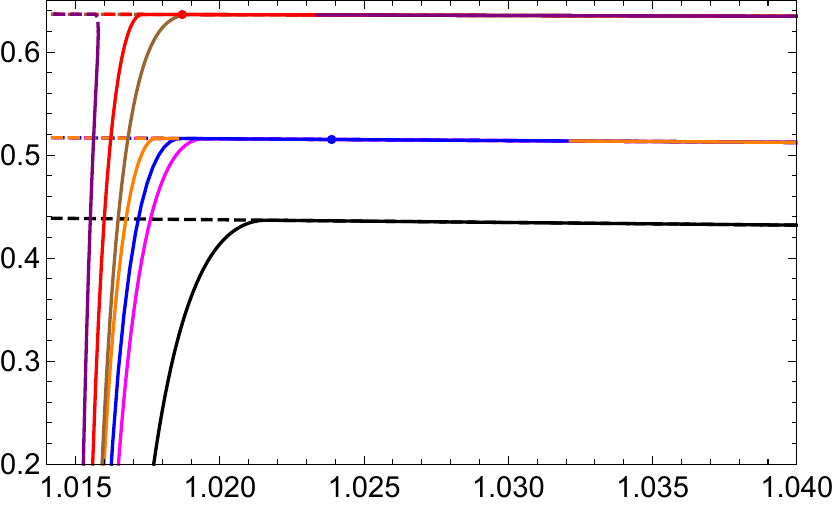}}
\end{overpic}\,\,
\put(5,20){\includegraphics[scale=0.75]{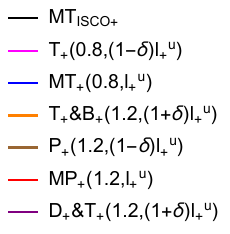}}
\caption{PEP $P(r_s)$ and MOB $z_\mob(r_s)$ for prograde $\M{MT}_{\isco+}$, (anti-)$\M{MP}_+$, $\M {MT}_+$, (anti-)$\M{P}_+$, $\M {T}_+$, $\M B_+$ and $\M D_+$ emitters with parameters $(E_s,l_s)$ for $a=0.9999$, where $\d=5\times10^{-4}$, $E_{\isco+}=0.618$ and the relevant values of $l^u_+(E_s)$ are given in Table~\ref{table:numericalLu2}. Solid curves are for ingoing ($s_r=-1$) emitters, while dashed curves are for outgoing ($s_r=+1$) emitters. The dots are for the results at the ISCO or the unstable double root $r_\ast$.}
\label{fig:highspin}
\end{figure}

In practice, we set black hole spin $a=0.9999$, and we consider the emitters on prograde orbits with angular momentum $l_s=(1\pm\d)l^u_+(E_s)$, where $\d=5\times10^{-4}<\epsilon$. The results of $P(r_s)$ and $z_{\mob}(r_s)$ for photon emissions from prograde orbits of different classes are shown in Fig.~\ref{fig:highspin}. Note that in this paper, the trapped emitters ($\M T$) inside the ISCO and the bounded emitters ($\M B$) with parts inside the ergosphere correspond to the bounded and deflected orbits in \cite{Yan:2021ygy}, respectively. The results in this paper include richer features than those in \cite{Yan:2021ygy}\footnote{The relevant results in this paper show the same behaviors as those in \cite{Yan:2021ygy} up to corrections of $z_{\mob,i}$ in the region $r_{p+}<r_s<r_{\isco+}$ (see Sec. \ref{sec:mob} for details).} where only the near-horizon region inside the ergosphere ($r_s\leq r_\isco<2$) were considered. We find that the case of ``plunging'' (including trapped) emitters from outside of the ISCO complements the plunging case discussed in \cite{Yan:2021ygy} and exhibits novel behaviors. Comparing with the marginal ``plunging'' emitters from the ISCO ($\M{MT}_{\isco}$), the other marginal emitters ($\M{MT}$ and $\M{MP}$) with larger angular momentum have lager PEP and MOB in the region outside the radii of the unstable circular orbits $r_\ast$. Therefore, high-energy emitters with near marginal angular momentum are more observable than the $\M{MT}_{\isco}$ emitter in a region inside the ISCO $r_s < r_\isco$. As $E_s$ is increased towards infinity, the double root $r_\ast$ moves towards the innermost photo orbit, that is $r_\ast \rightarrow r_{p+}$, and we find $P_{,i}(r_\ast)\rightarrow 1/2$ and $z_{\mob,i}(r_\ast)\rightarrow 1$. From the perspective of (near-)NHEK geometry, the above results mean that near marginal high-energy emitters would allow us to observe the very deep (near-NHEK) region of the near-horizon throat of a high-spin black hole.

\begin{table}[!tbh]
\centering
\caption{Several numerical values of $[E_s, l^u(E_s)]$ for $a=0.9999$, where $E_{\isco +}=0.618$.}
\begin{tabular}{c|c|c|c}
\hline\hline
$E_s$ & $E_{\isco +}$ & 0.8 & 1.2 \\
\hline
$l^{u}_+$ & 1.241 & 1.614 & 2.426 \\
\hline\hline
\end{tabular}
\label{table:numericalLu2}
\end{table}

\section{Conclusion}\label{sec:conclusion}
In this paper, we studied the observability of the emitters moving along the equatorial geodesics of a Kerr black hole with arbitrary spin $a$ by computing the PEP and MOB of the photons that escaped from these emitters. We considered the emitters with four basic motion classes: plunging, trapped, bounded, and deflected motions. The motion class of an emitter was determined by its energy $E_s$ and angular momentum $l_s$. In addition, an emitter's position and radial motion direction are labeled by $r_s$ and $s_r$, respectively. The results of PEP and MOB were shown in Figs.~\ref{fig:mtisco}--\ref{fig:highspin}, depending on the black hole spin $a$ and the emitter's parameters $(E_s,l_s,r_s,s_r)$.

We found that the results for the prograde and retrograde emitters with the same motion class exhibited similar features, as shown in Figs.~\ref{fig:mtisco}--\ref{fig:mptTogether}.
On the other hand, from Fig.~\ref{fig:ptbd}, we saw that the results for the emitters with different motion classes showed distinct features.
For photon emissions from a plunging or trapped emitter, as the emitter's radius is decreasing, the PEP decreases monotonously and reaches zero at the horizon, and the MOB is positive at the beginning but tends to $-\infty$ in the end.  For photon emissions from a bounded or deflected emitter, the PEP is always more than $1/2$ and the MOB is always positive, that is, the bounded and deflected emitters have good observability on their whole orbits. For photon emissions from a marginal plunging and trapped emitters in the region $r_s\geq r_\ast$, the PEP is also greater than $1/2$ and the MOB is positive as well, so that the marginal plunging and trapped emitters have good observability until they reach the position of the radial double root.
Therefore, for the nearly marginal emitters ($l_s\approx l^u$) with high energy, the observable region could extend to the places very close to the event horizon.

The results of this work could be of great relevance to the observability of the phenomena around an astrophysical black hole, including the image of an accretion disk \cite{EventHorizonTelescope:2019dse, EventHorizonTelescope:2022xnr,
Gralla:2019xty,Gralla:2019drh,Hou:2022gge,Hou:2022eev} and the signals of high-energy particle collisions \cite{Penrose:1969pc,Piran1975, Berti:2014lva,Schnittman:2014zsa,
Guo:2016vbt,Zhang:2016btg,Zhang:2020tfz}.
For example, a radiatively-inefficient accretion flow may consist of the plunging particles (perhaps near-critical high-energy plunging particles), thus our results suggest that one can truncate an accretion flow inside the ISCO when considering its appearance. In addition, a geometrically thick and optically thin disk may contain both accretion flow and outflow \cite{Yuan:2014gma}. Our results for the outgoing particles can be applied to the study of the appearance of the outflow.

In this study, we only focused on the equatorial emitters. It would be interesting to study the photon emissions from the emitters off the equatorial plane. We leave this work to the future.

\section*{Acknowledgments}
We thank Yan Liu and Jiang Long for the discussions on the classification of the Kerr geodesic orbits. The work is partly supported by NSFC Grant No. 12275004, 11775022, 11873044, 12205013 and 12305070. MG is also endorsed by ``the Fundamental Research Funds for the Central Universities'' with Grant No. 2021NTST13. HY is also supported by the Basic Research Program of Shanxi Province under Grant No. 202303021222018.

\bibliographystyle{utphys}
\bibliography{kerrnote}

\providecommand{\href}[2]{#2}\begingroup\raggedright\begin{thebibliography}{10}

\bibitem{EventHorizonTelescope:2019dse}
{\bfseries Event Horizon Telescope} Collaboration, K.~Akiyama {\em et~al.},
  ``{First M87 Event Horizon Telescope Results. I. The Shadow of the
  Supermassive Black Hole},''
  \href{http://dx.doi.org/10.3847/2041-8213/ab0ec7}{{\em Astrophys. J. Lett.}
  {\bfseries 875} (2019) L1}, \href{http://arxiv.org/abs/1906.11238}{{\ttfamily
  arXiv:1906.11238 [astro-ph.GA]}}.

\bibitem{EventHorizonTelescope:2022xnr}
{\bfseries Event Horizon Telescope} Collaboration, K.~Akiyama {\em et~al.},
  ``{First Sagittarius A* Event Horizon Telescope Results. I. The Shadow of the
  Supermassive Black Hole in the Center of the Milky Way},''
  \href{http://dx.doi.org/10.3847/2041-8213/ac6674}{{\em Astrophys. J. Lett.}
  {\bfseries 930} no.~2, (2022) L12}.

\bibitem{Cunha:2018acu}
P.~V.~P. Cunha and C.~A.~R. Herdeiro, ``{Shadows and strong gravitational
  lensing: a brief review},''
  \href{http://dx.doi.org/10.1007/s10714-018-2361-9}{{\em Gen. Rel. Grav.}
  {\bfseries 50} no.~4, (2018) 42},
  \href{http://arxiv.org/abs/1801.00860}{{\ttfamily arXiv:1801.00860 [gr-qc]}}.

\bibitem{Perlick:2021aok}
V.~Perlick and O.~Y. Tsupko, ``{Calculating black hole shadows: Review of
  analytical studies},''
  \href{http://dx.doi.org/10.1016/j.physrep.2021.10.004}{{\em Phys. Rept.}
  {\bfseries 947} (2022) 1--39},
  \href{http://arxiv.org/abs/2105.07101}{{\ttfamily arXiv:2105.07101 [gr-qc]}}.

\bibitem{Wei:2019pjf}
S.-W. Wei, Y.-C. Zou, Y.-X. Liu, and R.~B. Mann, ``{Curvature radius and Kerr
  black hole shadow},''
  \href{http://dx.doi.org/10.1088/1475-7516/2019/08/030}{{\em JCAP} {\bfseries
  08} (2019) 030}, \href{http://arxiv.org/abs/1904.07710}{{\ttfamily
  arXiv:1904.07710 [gr-qc]}}.

\bibitem{Banerjee:2022jog}
I.~Banerjee, S.~Chakraborty, and S.~SenGupta, ``{Hunting extra dimensions in
  the shadow of Sgr A*},''
  \href{http://dx.doi.org/10.1103/PhysRevD.106.084051}{{\em Phys. Rev. D}
  {\bfseries 106} no.~8, (2022) 084051},
  \href{http://arxiv.org/abs/2207.09003}{{\ttfamily arXiv:2207.09003 [gr-qc]}}.

\bibitem{Banerjee:2019nnj}
I.~Banerjee, S.~Chakraborty, and S.~SenGupta, ``{Silhouette of M87*: A New
  Window to Peek into the World of Hidden Dimensions},''
  \href{http://dx.doi.org/10.1103/PhysRevD.101.041301}{{\em Phys. Rev. D}
  {\bfseries 101} no.~4, (2020) 041301},
  \href{http://arxiv.org/abs/1909.09385}{{\ttfamily arXiv:1909.09385 [gr-qc]}}.

\bibitem{Mishra:2019trb}
A.~K. Mishra, S.~Chakraborty, and S.~Sarkar, ``{Understanding photon sphere and
  black hole shadow in dynamically evolving spacetimes},''
  \href{http://dx.doi.org/10.1103/PhysRevD.99.104080}{{\em Phys. Rev. D}
  {\bfseries 99} no.~10, (2019) 104080},
  \href{http://arxiv.org/abs/1903.06376}{{\ttfamily arXiv:1903.06376 [gr-qc]}}.

\bibitem{Li:2021btf}
Q.~Li, Y.~Zhu, and T.~Wang, ``{Gravitational effect of plasma particles on the
  shadow of Schwarzschild black holes},''
  \href{http://dx.doi.org/10.1140/epjc/s10052-021-09959-z}{{\em Eur. Phys. J.
  C} {\bfseries 82} no.~1, (2022) 2},
  \href{http://arxiv.org/abs/2102.00957}{{\ttfamily arXiv:2102.00957 [gr-qc]}}.

\bibitem{Wang:2022kvg}
M.~Wang, S.~Chen, and J.~Jing, ``{Chaotic shadows of black holes: a short
  review},'' \href{http://dx.doi.org/10.1088/1572-9494/ac6e5c}{{\em Commun.
  Theor. Phys.} {\bfseries 74} no.~9, (2022) 097401},
  \href{http://arxiv.org/abs/2205.05855}{{\ttfamily arXiv:2205.05855 [gr-qc]}}.

\bibitem{Li:2020drn}
P.-C. Li, M.~Guo, and B.~Chen, ``{Shadow of a Spinning Black Hole in an
  Expanding Universe},''
  \href{http://dx.doi.org/10.1103/PhysRevD.101.084041}{{\em Phys. Rev. D}
  {\bfseries 101} no.~8, (2020) 084041},
  \href{http://arxiv.org/abs/2001.04231}{{\ttfamily arXiv:2001.04231 [gr-qc]}}.

\bibitem{Zhong:2021mty}
Z.~Zhong, Z.~Hu, H.~Yan, M.~Guo, and B.~Chen, ``{QED effects on Kerr black hole
  shadows immersed in uniform magnetic fields},''
  \href{http://dx.doi.org/10.1103/PhysRevD.104.104028}{{\em Phys. Rev. D}
  {\bfseries 104} no.~10, (2021) 104028},
  \href{http://arxiv.org/abs/2108.06140}{{\ttfamily arXiv:2108.06140 [gr-qc]}}.

\bibitem{EventHorizonTelescope:2021bee}
{\bfseries Event Horizon Telescope} Collaboration, K.~Akiyama {\em et~al.},
  ``{First M87 Event Horizon Telescope Results. VII. Polarization of the
  Ring},'' \href{http://dx.doi.org/10.3847/2041-8213/abe71d}{{\em Astrophys. J.
  Lett.} {\bfseries 910} no.~1, (2021) L12},
  \href{http://arxiv.org/abs/2105.01169}{{\ttfamily arXiv:2105.01169
  [astro-ph.HE]}}.

\bibitem{Gralla:2019xty}
S.~E. Gralla, D.~E. Holz, and R.~M. Wald, ``{Black Hole Shadows, Photon Rings,
  and Lensing Rings},''
  \href{http://dx.doi.org/10.1103/PhysRevD.100.024018}{{\em Phys. Rev. D}
  {\bfseries 100} no.~2, (2019) 024018},
  \href{http://arxiv.org/abs/1906.00873}{{\ttfamily arXiv:1906.00873
  [astro-ph.HE]}}.

\bibitem{Himwich:2020msm}
E.~Himwich, M.~D. Johnson, A.~Lupsasca, and A.~Strominger, ``{Universal
  polarimetric signatures of the black hole photon ring},''
  \href{http://dx.doi.org/10.1103/PhysRevD.101.084020}{{\em Phys. Rev. D}
  {\bfseries 101} no.~8, (2020) 084020},
  \href{http://arxiv.org/abs/2001.08750}{{\ttfamily arXiv:2001.08750 [gr-qc]}}.

\bibitem{Johnson:2019ljv}
M.~D. Johnson {\em et~al.}, ``{Universal interferometric signatures of a black
  hole\textquoteright{}s photon ring},''
  \href{http://dx.doi.org/10.1126/sciadv.aaz1310}{{\em Sci. Adv.} {\bfseries 6}
  no.~12, (2020) eaaz1310}, \href{http://arxiv.org/abs/1907.04329}{{\ttfamily
  arXiv:1907.04329 [astro-ph.IM]}}.

\bibitem{Gralla:2020srx}
S.~E. Gralla, A.~Lupsasca, and D.~P. Marrone, ``{The shape of the black hole
  photon ring: A precise test of strong-field general relativity},''
  \href{http://dx.doi.org/10.1103/PhysRevD.102.124004}{{\em Phys. Rev. D}
  {\bfseries 102} no.~12, (2020) 124004},
  \href{http://arxiv.org/abs/2008.03879}{{\ttfamily arXiv:2008.03879 [gr-qc]}}.

\bibitem{Peng:2020wun}
J.~Peng, M.~Guo, and X.-H. Feng, ``{Influence of quantum correction on black
  hole shadows, photon rings, and lensing rings},''
  \href{http://dx.doi.org/10.1088/1674-1137/ac06bb}{{\em Chin. Phys. C}
  {\bfseries 45} no.~8, (2021) 085103},
  \href{http://arxiv.org/abs/2008.00657}{{\ttfamily arXiv:2008.00657 [gr-qc]}}.

\bibitem{Peng:2021osd}
J.~Peng, M.~Guo, and X.-H. Feng, ``{Observational signature and additional
  photon rings of an asymmetric thin-shell wormhole},''
  \href{http://dx.doi.org/10.1103/PhysRevD.104.124010}{{\em Phys. Rev. D}
  {\bfseries 104} no.~12, (2021) 124010},
  \href{http://arxiv.org/abs/2102.05488}{{\ttfamily arXiv:2102.05488 [gr-qc]}}.

\bibitem{Chen:2022nbb}
Y.~Chen, R.~Roy, S.~Vagnozzi, and L.~Visinelli, ``{Superradiant evolution of
  the shadow and photon ring of Sgr A\ensuremath{\star}},''
  \href{http://dx.doi.org/10.1103/PhysRevD.106.043021}{{\em Phys. Rev. D}
  {\bfseries 106} no.~4, (2022) 043021},
  \href{http://arxiv.org/abs/2205.06238}{{\ttfamily arXiv:2205.06238
  [astro-ph.HE]}}.

\bibitem{Gralla:2017ufe}
S.~E. Gralla, A.~Lupsasca, and A.~Strominger, ``{Observational Signature of
  High Spin at the Event Horizon Telescope},''
  \href{http://dx.doi.org/10.1093/mnras/sty039}{{\em Mon. Not. Roy. Astron.
  Soc.} {\bfseries 475} no.~3, (2018) 3829--3853},
  \href{http://arxiv.org/abs/1710.11112}{{\ttfamily arXiv:1710.11112
  [astro-ph.HE]}}.

\bibitem{Guo:2018kis}
M.~Guo, N.~A. Obers, and H.~Yan, ``{Observational signatures of near-extremal
  Kerr-like black holes in a modified gravity theory at the Event Horizon
  Telescope},'' \href{http://dx.doi.org/10.1103/PhysRevD.98.084063}{{\em Phys.
  Rev. D} {\bfseries 98} no.~8, (2018) 084063},
  \href{http://arxiv.org/abs/1806.05249}{{\ttfamily arXiv:1806.05249 [gr-qc]}}.

\bibitem{Yan:2019etp}
H.~Yan, ``{Influence of a plasma on the observational signature of a high-spin
  Kerr black hole},'' \href{http://dx.doi.org/10.1103/PhysRevD.99.084050}{{\em
  Phys. Rev. D} {\bfseries 99} no.~8, (2019) 084050},
  \href{http://arxiv.org/abs/1903.04382}{{\ttfamily arXiv:1903.04382 [gr-qc]}}.

\bibitem{Guo:2019lur}
M.~Guo, S.~Song, and H.~Yan, ``{Observational signature of a near-extremal
  Kerr-Sen black hole in the heterotic string theory},''
  \href{http://dx.doi.org/10.1103/PhysRevD.101.024055}{{\em Phys. Rev. D}
  {\bfseries 101} no.~2, (2020) 024055},
  \href{http://arxiv.org/abs/1911.04796}{{\ttfamily arXiv:1911.04796 [gr-qc]}}.

\bibitem{EventHorizonTelescope:2021srq}
{\bfseries Event Horizon Telescope} Collaboration, K.~Akiyama {\em et~al.},
  ``{First M87 Event Horizon Telescope Results. VIII. Magnetic Field Structure
  near The Event Horizon},''
  \href{http://dx.doi.org/10.3847/2041-8213/abe4de}{{\em Astrophys. J. Lett.}
  {\bfseries 910} no.~1, (2021) L13},
  \href{http://arxiv.org/abs/2105.01173}{{\ttfamily arXiv:2105.01173
  [astro-ph.HE]}}.

\bibitem{EventHorizonTelescope:2021btj}
{\bfseries Event Horizon Telescope} Collaboration, K.~Akiyama {\em et~al.},
  ``{The Polarized Image of a Synchrotron-emitting Ring of Gas Orbiting a Black
  Hole},'' \href{http://dx.doi.org/10.3847/1538-4357/abf117}{{\em Astrophys.
  J.} {\bfseries 912} no.~1, (2021) 35},
  \href{http://arxiv.org/abs/2105.01804}{{\ttfamily arXiv:2105.01804
  [astro-ph.HE]}}.

\bibitem{Junior:2021dyw}
H.~C. D.~L. Junior, P.~V.~P. Cunha, C.~A.~R. Herdeiro, and L.~C.~B. Crispino,
  ``{Shadows and lensing of black holes immersed in strong magnetic fields},''
  \href{http://dx.doi.org/10.1103/PhysRevD.104.044018}{{\em Phys. Rev. D}
  {\bfseries 104} no.~4, (2021) 044018},
  \href{http://arxiv.org/abs/2104.09577}{{\ttfamily arXiv:2104.09577 [gr-qc]}}.

\bibitem{Hou:2022eev}
Y.~Hou, Z.~Zhang, H.~Yan, M.~Guo, and B.~Chen, ``{Image of a Kerr-Melvin black
  hole with a thin accretion disk},''
  \href{http://dx.doi.org/10.1103/PhysRevD.106.064058}{{\em Phys. Rev. D}
  {\bfseries 106} no.~6, (2022) 064058},
  \href{http://arxiv.org/abs/2206.13744}{{\ttfamily arXiv:2206.13744 [gr-qc]}}.

\bibitem{Hu:2022sej}
Z.~Hu, Y.~Hou, H.~Yan, M.~Guo, and B.~Chen, ``{Polarized images of synchrotron
  radiations in curved spacetime},''
  \href{http://dx.doi.org/10.1140/epjc/s10052-022-11144-9}{{\em Eur. Phys. J.
  C} {\bfseries 82} no.~12, (2022) 1166},
  \href{http://arxiv.org/abs/2203.02908}{{\ttfamily arXiv:2203.02908 [gr-qc]}}.

\bibitem{Lee:2022rtg}
T.-C. Lee, Z.~Hu, M.~Guo, and B.~Chen, ``{Circular orbits and polarized images
  of charged particles orbiting a Kerr black hole with a weak magnetic
  field},'' \href{http://dx.doi.org/10.1103/PhysRevD.108.024008}{{\em Phys.
  Rev. D} {\bfseries 108} no.~2, (2023) 024008},
  \href{http://arxiv.org/abs/2211.04143}{{\ttfamily arXiv:2211.04143 [gr-qc]}}.

\bibitem{Synge:1966okc}
J.~L. Synge, ``{The Escape of Photons from Gravitationally Intense Stars},''
  \href{http://dx.doi.org/10.1093/mnras/131.3.463}{{\em Mon. Not. Roy. Astron.
  Soc.} {\bfseries 131} no.~3, (1966) 463--466}.

\bibitem{etde_271389}
O.~Semerak, ``Photon escape cones in the kerr field,''
  \href{http://dx.doi.org/10.5169/seals-116907}{{\em Helv. Phys. Acta}
  {\bfseries 69} no.~1, (1996) 69--80}.

\bibitem{Stuchlik:2018qyz}
Z.~Stuchl\'\i{}k, D.~Charbul\'ak, and J.~Schee, ``{Light escape cones in local
  reference frames of Kerr\textendash{}de Sitter black hole spacetimes and
  related black hole shadows},''
  \href{http://dx.doi.org/10.1140/epjc/s10052-018-5578-6}{{\em Eur. Phys. J. C}
  {\bfseries 78} no.~3, (2018) 180},
  \href{http://arxiv.org/abs/1811.00072}{{\ttfamily arXiv:1811.00072 [gr-qc]}}.

\bibitem{Ogasawara:2019mir}
K.~Ogasawara, T.~Igata, T.~Harada, and U.~Miyamoto, ``{Escape probability of a
  photon emitted near the black hole horizon},''
  \href{http://dx.doi.org/10.1103/PhysRevD.101.044023}{{\em Phys. Rev. D}
  {\bfseries 101} no.~4, (2020) 044023},
  \href{http://arxiv.org/abs/1910.01528}{{\ttfamily arXiv:1910.01528 [gr-qc]}}.

\bibitem{Zhang:2020pay}
M.~Zhang and J.~Jiang, ``{Emissions of photons near the horizons of Kerr-Sen
  black holes},'' \href{http://dx.doi.org/10.1103/PhysRevD.102.124012}{{\em
  Phys. Rev. D} {\bfseries 102} no.~12, (2020) 124012},
  \href{http://arxiv.org/abs/2004.11087}{{\ttfamily arXiv:2004.11087 [gr-qc]}}.

\bibitem{Igata:2019hkz}
T.~Igata, K.~Nakashi, and K.~Ogasawara, ``{Observability of the innermost
  stable circular orbit in a near-extremal Kerr black hole},''
  \href{http://dx.doi.org/10.1103/PhysRevD.101.044044}{{\em Phys. Rev. D}
  {\bfseries 101} no.~4, (2020) 044044},
  \href{http://arxiv.org/abs/1910.12682}{{\ttfamily arXiv:1910.12682
  [astro-ph.HE]}}.

\bibitem{Gates:2020els}
D.~E.~A. Gates, S.~Hadar, and A.~Lupsasca, ``{Photon emission from circular
  equatorial Kerr orbiters},''
  \href{http://dx.doi.org/10.1103/PhysRevD.103.044050}{{\em Phys. Rev. D}
  {\bfseries 103} no.~4, (2021) 044050},
  \href{http://arxiv.org/abs/2010.07330}{{\ttfamily arXiv:2010.07330 [gr-qc]}}.

\bibitem{Igata:2021njn}
T.~Igata, K.~Kohri, and K.~Ogasawara, ``{Photon emission from inside the
  innermost stable circular orbit},''
  \href{http://dx.doi.org/10.1103/PhysRevD.103.104028}{{\em Phys. Rev. D}
  {\bfseries 103} no.~10, (2021) 104028},
  \href{http://arxiv.org/abs/2102.13427}{{\ttfamily arXiv:2102.13427 [gr-qc]}}.

\bibitem{Ogasawara:2020frt}
K.~Ogasawara and T.~Igata, ``{Complete classification of photon escape in the
  Kerr black hole spacetime},''
  \href{http://dx.doi.org/10.1103/PhysRevD.103.044029}{{\em Phys. Rev. D}
  {\bfseries 103} no.~4, (2021) 044029},
  \href{http://arxiv.org/abs/2011.04380}{{\ttfamily arXiv:2011.04380 [gr-qc]}}.

\bibitem{Ogasawara:2021yfe}
K.~Ogasawara and T.~Igata, ``{Photon escape in the extremal Kerr black hole
  spacetime},'' \href{http://dx.doi.org/10.1103/PhysRevD.105.024031}{{\em Phys.
  Rev. D} {\bfseries 105} no.~2, (2022) 024031},
  \href{http://arxiv.org/abs/2111.03243}{{\ttfamily arXiv:2111.03243 [gr-qc]}}.

\bibitem{Bardeen:1999px}
J.~M. Bardeen and G.~T. Horowitz, ``{The Extreme Kerr throat geometry: A Vacuum
  analog of AdS(2) x S**2},''
  \href{http://dx.doi.org/10.1103/PhysRevD.60.104030}{{\em Phys. Rev. D}
  {\bfseries 60} (1999) 104030},
  \href{http://arxiv.org/abs/hep-th/9905099}{{\ttfamily arXiv:hep-th/9905099}}.

\bibitem{Bredberg:2009pv}
I.~Bredberg, T.~Hartman, W.~Song, and A.~Strominger, ``{Black Hole
  Superradiance From Kerr/CFT},''
  \href{http://dx.doi.org/10.1007/JHEP04(2010)019}{{\em JHEP} {\bfseries 04}
  (2010) 019}, \href{http://arxiv.org/abs/0907.3477}{{\ttfamily arXiv:0907.3477
  [hep-th]}}.

\bibitem{Gralla:2015rpa}
S.~E. Gralla, A.~P. Porfyriadis, and N.~Warburton, ``{Particle on the Innermost
  Stable Circular Orbit of a Rapidly Spinning Black Hole},''
  \href{http://dx.doi.org/10.1103/PhysRevD.92.064029}{{\em Phys. Rev. D}
  {\bfseries 92} no.~6, (2015) 064029},
  \href{http://arxiv.org/abs/1506.08496}{{\ttfamily arXiv:1506.08496 [gr-qc]}}.

\bibitem{Yan:2021yuo}
H.~Yan, M.~Guo, and B.~Chen, ``{Observability of zero-angular-momentum sources
  near Kerr black holes},''
  \href{http://dx.doi.org/10.1140/epjc/s10052-021-09649-w}{{\em Eur. Phys. J.
  C} {\bfseries 81} no.~9, (2021) 847},
  \href{http://arxiv.org/abs/2104.07889}{{\ttfamily arXiv:2104.07889 [gr-qc]}}.

\bibitem{Yan:2021ygy}
H.~Yan, Z.~Hu, M.~Guo, and B.~Chen, ``{Photon emissions from near-horizon
  extremal and near-extremal Kerr equatorial emitters},''
  \href{http://dx.doi.org/10.1103/PhysRevD.104.124005}{{\em Phys. Rev. D}
  {\bfseries 104} no.~12, (2021) 124005},
  \href{http://arxiv.org/abs/2108.09051}{{\ttfamily arXiv:2108.09051 [gr-qc]}}.

\bibitem{Bardeen:1973tla}
J.~M. Bardeen, ``{Timelike and null geodesics in the Kerr metric},'' in {\em
  {Les Houches Summer School of Theoretical Physics}: {Black Holes}}.
\newblock 1973.

\bibitem{Compere:2021bkk}
G.~Comp\`ere, Y.~Liu, and J.~Long, ``{Classification of radial Kerr geodesic
  motion},'' \href{http://dx.doi.org/10.1103/PhysRevD.105.024075}{{\em Phys.
  Rev. D} {\bfseries 105} no.~2, (2022) 024075},
  \href{http://arxiv.org/abs/2106.03141}{{\ttfamily arXiv:2106.03141 [gr-qc]}}.

\bibitem{Bardeen:1972fi}
J.~M. Bardeen, W.~H. Press, and S.~A. Teukolsky, ``{Rotating black holes:
  Locally nonrotating frames, energy extraction, and scalar synchrotron
  radiation},'' \href{http://dx.doi.org/10.1086/151796}{{\em Astrophys. J.}
  {\bfseries 178} (1972) 347}.

\bibitem{Gralla:2019drh}
S.~E. Gralla and A.~Lupsasca, ``{Lensing by Kerr Black Holes},''
  \href{http://dx.doi.org/10.1103/PhysRevD.101.044031}{{\em Phys. Rev. D}
  {\bfseries 101} no.~4, (2020) 044031},
  \href{http://arxiv.org/abs/1910.12873}{{\ttfamily arXiv:1910.12873 [gr-qc]}}.

\bibitem{Hou:2022gge}
Y.~Hou, P.~Liu, M.~Guo, H.~Yan, and B.~Chen, ``{Multi-level images around
  Kerr\textendash{}Newman black holes},''
  \href{http://dx.doi.org/10.1088/1361-6382/ac8860}{{\em Class. Quant. Grav.}
  {\bfseries 39} no.~19, (2022) 194001},
  \href{http://arxiv.org/abs/2203.02755}{{\ttfamily arXiv:2203.02755 [gr-qc]}}.

\bibitem{Penrose:1969pc}
R.~Penrose, ``{Gravitational collapse: The role of general relativity},''
  \href{http://dx.doi.org/10.1023/A:1016578408204}{{\em Riv. Nuovo Cim.}
  {\bfseries 1} (1969) 252--276}.

\bibitem{Piran1975}
J.~K. T.~Piran, J.~Shaham, ``High efficiency of the penrose mechanism for
  particle collisions,'' {\em Astrophys.J.} {\bfseries 196} (1975) L107.

\bibitem{Berti:2014lva}
E.~Berti, R.~Brito, and V.~Cardoso, ``{Ultrahigh-energy debris from the
  collisional Penrose process},''
  \href{http://dx.doi.org/10.1103/PhysRevLett.114.251103}{{\em Phys. Rev.
  Lett.} {\bfseries 114} no.~25, (2015) 251103},
  \href{http://arxiv.org/abs/1410.8534}{{\ttfamily arXiv:1410.8534 [gr-qc]}}.

\bibitem{Schnittman:2014zsa}
J.~D. Schnittman, ``{Revised upper limit to energy extraction from a Kerr black
  hole},'' \href{http://dx.doi.org/10.1103/PhysRevLett.113.261102}{{\em Phys.
  Rev. Lett.} {\bfseries 113} (2014) 261102},
  \href{http://arxiv.org/abs/1410.6446}{{\ttfamily arXiv:1410.6446
  [astro-ph.HE]}}.

\bibitem{Guo:2016vbt}
M.~Guo and S.~Gao, ``{Kerr black holes as accelerators of spinning test
  particles},'' \href{http://dx.doi.org/10.1103/PhysRevD.93.084025}{{\em Phys.
  Rev. D} {\bfseries 93} no.~8, (2016) 084025},
  \href{http://arxiv.org/abs/1602.08679}{{\ttfamily arXiv:1602.08679 [gr-qc]}}.

\bibitem{Zhang:2016btg}
Y.-P. Zhang, B.-M. Gu, S.-W. Wei, J.~Yang, and Y.-X. Liu, ``{Charged spinning
  black holes as accelerators of spinning particles},''
  \href{http://dx.doi.org/10.1103/PhysRevD.94.124017}{{\em Phys. Rev. D}
  {\bfseries 94} no.~12, (2016) 124017},
  \href{http://arxiv.org/abs/1608.08705}{{\ttfamily arXiv:1608.08705 [gr-qc]}}.

\bibitem{Zhang:2020tfz}
M.~Zhang and J.~Jiang, ``{Escape probability of particle from Kerr-Sen black
  hole},'' \href{http://dx.doi.org/10.1016/j.nuclphysb.2021.115313}{{\em Nucl.
  Phys. B} {\bfseries 964} (2021) 115313},
  \href{http://arxiv.org/abs/2004.03367}{{\ttfamily arXiv:2004.03367 [gr-qc]}}.

\bibitem{Yuan:2014gma}
F.~Yuan and R.~Narayan, ``{Hot Accretion Flows Around Black Holes},''
  \href{http://dx.doi.org/10.1146/annurev-astro-082812-141003}{{\em Ann. Rev.
  Astron. Astrophys.} {\bfseries 52} (2014) 529--588},
  \href{http://arxiv.org/abs/1401.0586}{{\ttfamily arXiv:1401.0586
  [astro-ph.HE]}}.

\end{thebibliography}\endgroup

\end{document}